\documentclass[sigconf]{acmart}

\usepackage{xcolor}
\usepackage{subcaption}
\usepackage{booktabs} \usepackage{tikz}

\usepackage{todonotes}

\usepackage{algorithm,graphicx}
\usepackage[noend]{algpseudocode}
\usepackage{enumitem}
\usepackage{makecell}
\usepackage{algpseudocode}
\algnewcommand{\LineComment}[1]{\Statex \(\triangleright\) #1}
\makeatletter
\newcommand{\algmargin}{\the\ALG@thistlm}
\makeatletter
\newcommand*{\rom}[1]{\expandafter\@slowromancap\romannumeral #1@}
\makeatother
\newlength{\whilewidth}
\algdef{SE}[parWHILE]{parWhile}{EndparWhile}[1]
  {\parbox[t]{\dimexpr\linewidth-\algmargin}{     \hangindent\whilewidth\strut\algorithmicwhile\ #1\ \algorithmicdo\strut}}{\algorithmicend\ \algorithmicwhile}\algnewcommand{\parState}[1]{\State  \parbox[t]{\dimexpr\linewidth-\algmargin}{\strut #1\strut}}
\algnewcommand{\parRequire}[1]{\Require  \parbox[t]{\dimexpr\linewidth-\algmargin}{\strut #1\strut}}

\usepackage{hyperref}
\usepackage{cleveref}
\usepackage{etoolbox}

\newtheorem{definition}{Definition}
\newtheorem{theorem}{Theorem}

\newtheorem{lemma}{Lemma}

\crefname{section}{Section}{Sections}
\crefname{subsection}{Section}{Sections}
\crefname{definition}{Definition}{Definitions}
\crefname{proposition}{Proposition}{Propositions}
\crefname{lemma}{Lemma}{Lemmas}
\crefname{theorem}{Theorem}{Theorems}
\crefname{corollary}{Corollary}{Corollaries}
\crefname{example}{Example}{Examples}
\crefname{figure}{Figure}{Figures}
\crefname{assumption}{Assumption}{Assumptions}
\crefname{remark}{Remark}{Remarks}
\crefname{running}{Running Example}{Running Examples}
\crefname{algorithm}{Algorithm}{Algorithms}

\def\delequal{\mathrel{\ensurestackMath{\stackon[1pt]{=}{\scriptstyle\Delta}}}}

\newcommand\old[1]{{\color{gray} #1}}

\renewcommand\old[1]{}

\setcopyright{rightsretained}
\usepackage{stackengine}

\acmDOI{10.475/123_4}

\acmISBN{123-4567-24-567/08/06}

\acmConference[ICCPS'22]{International Conference on Cyber-Physical Systems}{May 2022}{Milan, Italy}
\acmYear{2022}
\copyrightyear{2022}

\renewcommand\footnotetextcopyrightpermission[1]{}

\begin{document}
\title{Learning-Based Vulnerability Analysis of \\Cyber-Physical Systems}

\author{Amir Khazraei}
\affiliation{%
 \institution{Duke University}
 \city{Durham}
 \state{NC}$  $
 \postcode{27708}
}
\email{amir.khazraei@duke.edu}

\author{Spencer Hallyburton}
\affiliation{%
 \institution{Duke University}
 \city{Durham}
 \state{NC}$  $
 \postcode{27708}
 }
\email{spencer.hallyburton@duke.edu}

\author{Qitong Gao}
\affiliation{%
 \institution{Duke University}
 \city{Durham}
 \state{NC}$  $
 \postcode{27708}
 }
\email{qitong.gao@duke.edu}

\author{Yu Wang}
\affiliation{%
 \institution{University of Florida}
\city{Gainesville}
 \state{FL}$  $
 \postcode{32611}
}
\email{yuwang1@ufl.edu}

\author{Miroslav Pajic}
\affiliation{%
 \institution{Duke University}
 \city{Durham}
 \state{NC}$  $
 \postcode{27708}
}
\email{miroslav.pajic@duke.edu}
\thanks{This work is sponsored in part by the ONR under agreements N00014-17-1-2504 and N00014-20-1-2745, AFOSR under award number FA9550-19-1-0169, NSF under CNS-1652544 award as well as the National AI Institute for Edge Computing Leveraging Next Generation Wireless Networks, Grant CNS-2112562, and a grant from~Intel.} 

\begin{abstract}

This work focuses on the use of deep learning for vulnerability analysis of cyber-physical systems (CPS). Specifically, we consider a control architecture widely used in CPS, 
where the low-level control is based on a feedback controller and an observer (e.g., the extended Kalman filter (EKF)), while also employing an anomaly detector. To facilitate analyzing the impact potential sensing attacks could have on systems with \textit{general nonlinear} dynamics, we develop \textit{learning-enabled attack generators} capable of designing stealthy attacks that
maximally degrade system operation. 
We show how such problem can be cast within a learning-based \textit{grey-box} framework where only 
parts of the runtime information are known to the attacker. We then introduce two 
methods for generating \textit{effective stealthy attacks}, 
based on feed-forward neural networks~(FNN) and recurrent neural networks (RNN). 
Both types of {attack-generator models} are trained offline, using a cost function that combines the attack impact on the estimation error (and thus control) and the residual signal used for anomaly detection; this enables the trained models 
to recursively generate  effective yet stealthy sensor attacks in real-time while requiring different levels of system information at \textit{runtime}. The effectiveness of the proposed methods is demonstrated  on several case studies with varying levels of complexity and nonlinearity: inverted pendulum, autonomous driving vehicles (ADV), and unmanned areal vehicles (UAVs).

\end{abstract}

\keywords{Security of Cyber-Physical Systems, Deep Learning, Stealthy Attacks, Vulnerability Analysis, Secure Autonomy}

\maketitle
\section{Introduction}  
\label{sec:intro} 
Although many cyber-physical systems (CPS) operate in safety-critical scenarios and the heterogeneous component connectivity provides numerous possible points of attack, most of existing  systems 
are only weakly protected by legacy components, such as anomaly detectors. 
The challenge~of~securing CPS is even more formidable 
as the long system lifetime and resource constraints prevent the full use of new and existing security mechanisms. On the other hand, 
security-aware resource allocation can significantly reduce the security-related overhead and thus system cost~\cite{lesi_tecs17,lesi_rtss17,lesi_tcps20}; the idea is to focus on protecting the critical system components and communication links, which if compromised could significantly degrade system performance. Yet, to achieve this, 
we need methods to  analyze system vulnerability, in terms of performance degradation under attack, for different types of attacks. 

In this work, we investigate the use of deep learning for the vulnerability analysis of control mechanisms in CPS, focusing on attacks on system sensing. 
CPS controllers are commonly equipped with a state estimator used for low-level control and anomaly detection. Therefore, attacks on sensing may have tremendous impact on the system performance (i.e., quality of control -- QoC), by introducing errors in state estimation. In such setting, to maximize the damage by exploiting the compromised components, the goal of the attacker is to modify sensor measurements delivered to the controller such that the system is forced into an unsafe region, while the attack remains undetected (i.e., stealthy). 

Consequently, a critical part of the vulnerability analysis are methods/models for design of effective and stealthy attack vectors. Such \textit{attack generators} should capture how both attack stealthiness and effectiveness are affected by system dynamics, which in 
general is nonlinear; this prevents the use of existing model-based methods derived for linear time-invariant (LTI) systems (e.g., \cite{mo2010false,kwon2014analysis,smith2015covert,teixeira2012revealing,sui2020vulnerability}).
To address this challenge, we employ deep learning to develop  generators of such 
effective yet stealthy attack signals (i.e., time series).  
Specifically, we provide grey-box yet model-free methods that only use the estimator model (and not the plant model) to train stealthy attack generators. 
{We show that to remain stealthy, the attack generator should exhibit a suitable unstable dynamics, resulting \emph{{in large (potentially unbounded) attack vectors over time}}; this also prevents the use of standard robustness-based analysis techniques that consider performance degradation in the presence of {{bounded}} input disturbances.}

We propose two attack-generator models for design of such stealthy attack vectors. Each model requires different levels of runtime information from the state estimator -- i.e., the current sensor measurements and the previous state estimation, or only the current sensor measurements. The two models, based on feed-forward neural networks (FNN) and recurrent neural networks (RNN), are trained offline using a cost function that captures the impact the attack would have on the estimation error (and thus QoC) as well as  stealthiness requirements. To capture the expectation operation in the cost function, we employ Monte Carlo (MC) simulation.  

Finally, we illustrate the use and evaluate effectiveness of our approach on three case studies, from inverted pendulum to autonomous driving vehicles (ADVs) and unmanned aerial vehicles (UAVs). 
%
We show that when a suitably large time-duration is used for offline training, on average the learned FNN-based attacks  slightly outperform the RNN-based attacks. However, this comes at a price;~they require additional system information at the runtime -- i.e., local state estimation.
Furthermore, we demonstrate attack generalizability on a complex case study based on CARLA urban autonomous driving simulator~\cite{dosovitskiy2017carla}; by training the attack generator models on a simple path and then showing their effectiveness on more complex scenarios. We also show the 
robustness of the proposed attack design 
to changes in 
sensing frequencies.

\subsubsection*{Related Work}
This work is related in spirit to adversarial machine learning focused on methods to generate adversarial examples that degrade performance of deep neural network (DNN) models. The initial work~\cite{szegedy2013intriguing} showed that even small perturbations of a DNN's input could drastically change the output, 
starting a line of research 
on vulnerability (in terms of robustness) 
of DNNs. For instance, \cite{szegedy2013intriguing,goodfellow2014explaining,papernot2017practical,yuan2019adversarial} study vulnerability of classifiers by adding a small perturbation $z$ to the input $x$, and design an adversarial example $x^*=x+z$ that results in miss-classification error $C(x^*)\neq C(x)$, for some classifier $C$. In \cite{cao2019adversarial, sun2020towards, hallyburton2021security}, the same idea is applied to self-driving vehicles, where the attacker's goal is to fool a DNN perception model into `detecting' fake objects in front of the vehicle 
or removing an existing object, in order to maliciously alter its driving decisions. 
Some recent works also consider adversarial machine learning beyond the image domain~\cite{li_ConAMLConstrainedAdversarial_2020,zizzo_AdversarialMachineLearning_2019,feng_DeepLearningbasedFramework_2017}. For example, \cite{li_ConAMLConstrainedAdversarial_2020} studies vulnerability of machine learning models applied in CPS by proposing methods for generating adversarial examples that satisfy some physical constraints. 

However, the common assumption among such approaches  (e.g., \cite{goodfellow2014explaining,kurakin2016adversarial,moosavi2016deepfool,croce2020minimally,carlini2017towards,sun2020towards,li_ConAMLConstrainedAdversarial_2020,zizzo_AdversarialMachineLearning_2019,feng_DeepLearningbasedFramework_2017}) 
is 
that the predicted target only depends on its input and not internal dynamics -- i.e., previous states; {thus, considering bounded perturbation and a single time-instance (i.e., without longitudinal effects) was sufficient in those cases.} On the other hand, to address requirements of attacking a system with internal dynamics, 
in this work, we show that we \emph{have to consider attack models whose output should also depend on the previous outputs (i.e., previous state)}. In addition, unlike in the aforementioned works, due to the CPS control perspective, both input and outputs belong to a continuous space.

From the control perspective, it was shown that deep reinforcement learning models are susceptible to adversarial examples \cite{Weng2020Toward,lin2017tactics,huang2017adversarial}. Since the mapping between the observation to actions is achieved by a DNN, the idea has been to add  small {(i.e., bounded)} perturbations 
on observations to alter the actions in a way that minimizes the expected cumulative reward function, 
even driving the system to unsafe states~\cite{Weng2020Toward}. 
{On the other hand, in this work, we show that due to the stealthiness constraint, the time series for an effective additive attacks (on sensor measurements) cannot be bounded; rather, the injected attack signal over time should comply with a certain underlying unstable dynamics that depends on the dynamics of the controlled physical process.}

Finally, significant 
efforts focused on model-based (i.e., using more traditional control techniques) design of effective stealthy attacks on CPS controllers \cite{mo2010false,smith2015covert,teixeira2012revealing,sui2020vulnerability,khazraei_acc20,khazraei2017new,liu2011false,kwon2014analysis}, 
including replay~\cite{mo2009secure}, covert~\cite{smith2015covert}, zero dynamic~\cite{teixeira2012revealing} and false data injection attacks~\cite{mo2010false}. However, these methods can be used only for 
LTI dynamical systems, and thus have limited applicability in practice. 
For example, \cite{shen2020drift} designs stealthy sensing attacks 
on autonomous vehicles. However, the work 
effectively employs a standard 
LTI-based attack design where the attack vector is obtained using evolution of the system's linearized model around the equilibrium point. In this work, we show that, as expected, such LTI-based attack designs are only effective in a small neighborhood around the equilibrium point where the linear approximation is valid. As the states move further away from the equilibrium point, the error of the linear approximation 
significantly increases, resulting in attack detection. 

Some recent works have also studied learning-based attack design for control systems~\cite{rangi2021learning,khojasteh2020learning}. However, they assume the system has an LTI dynamical model as opposed to this work where we design stealthy impactful attacks for general nonlinear systems. 
To the best of our knowledge, this is the first study focused on design of stealthy attack signals -- potentially unbounded (in size) vector time-series -- that degrade QoC performance of 
control systems with general nonlinear dynamics, and for which only limited knowledge of the physical model is available. 

\subsubsection*{Notation}
\label{sec:notation}

We use 
$\mathbb{R}$ to denote the set of real numbers, and $\mathbb{P}$ and $\mathbb{E}$ denote the probability and expectation for a random variable.  For a matrix $A$, $A^T$ denotes its transpose and for a square matrix, $trace(A)$ denotes its trace. 
In addition, $I$ is the identity matrix in general, while $I_p$ is the identity matrix with dimension $p\times p$. 
Matrix $A\in \mathbb{R}^{n\times n}$ is positive semideﬁnite (denoted by $A\succeq 0$) if $x^TAx\geq 0$ holds for all $x\in \mathbb{R}^n$. 
For a vector $x\in{\mathbb{R}^n}$, $||x||_p$ is the $p$-norm of~$x$; when $p$ is not specified, the 2-norm is implied. 
Also, $\text{supp}(x)$ denotes the indices of the nonzero elements of $x\in\mathbb{R}^n$ -- i.e., $\text{supp}(x)=\{i~|~i\in\{1,...,n\}, x_i\neq0\}$. 
Finally, a function $h:\mathbb{R}^{n}\to \mathbb{R}^{p}$ is L-Lipschitz if for any $x,y\in \mathbb{R}^{n}$ it holds that $||h(x)-h(y)||\leq L ||x-y||$.

\section{System and Attack Models} \label{sec:back} 

\begin{figure}
\includegraphics[width=0.502\textwidth]{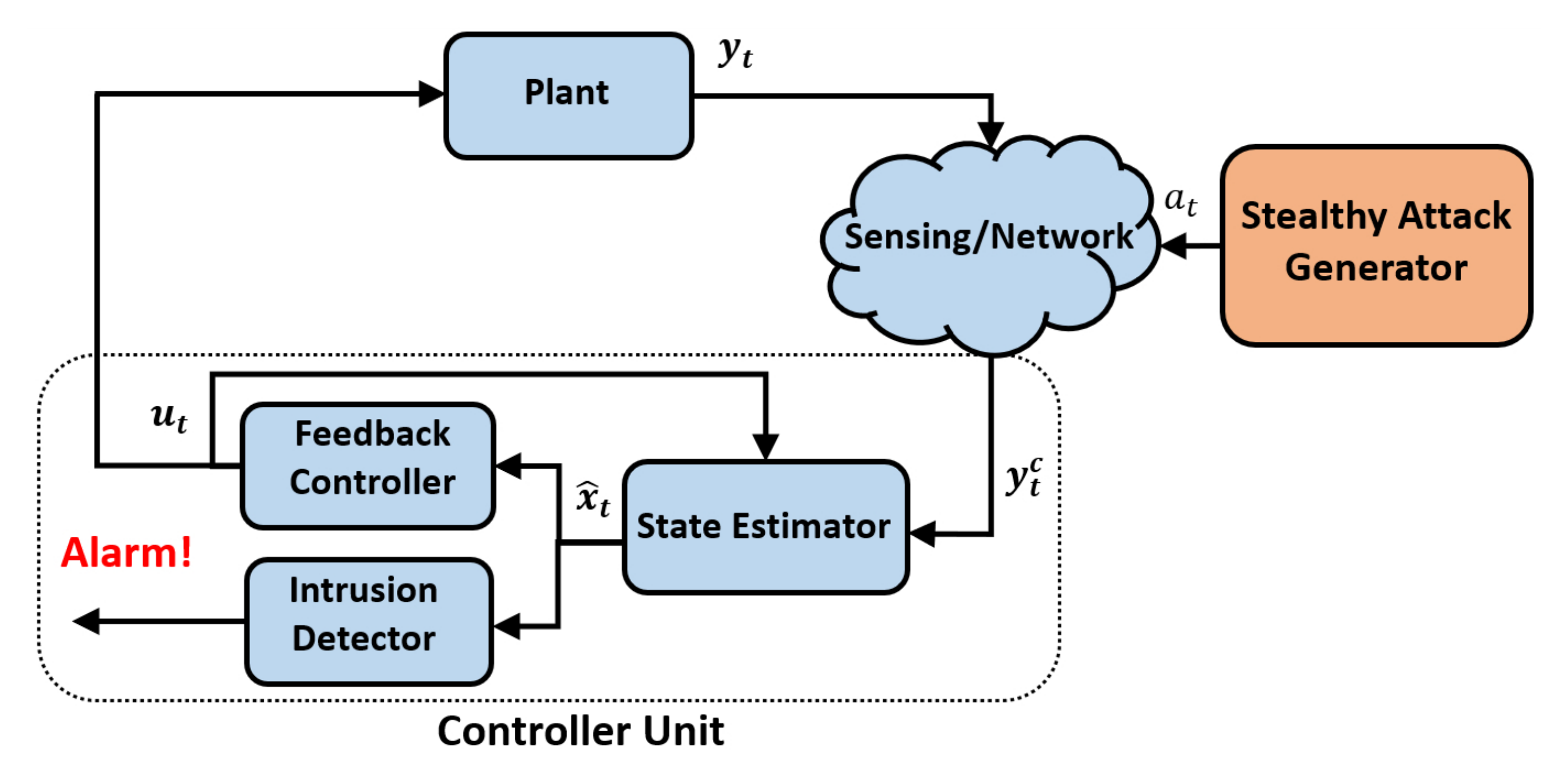}
\vspace{-20pt}
\caption{CPS architecture under attacks on system sensing; the considered general attack model captures the impact of both network-based attacks (e.g., man-in-the-middle attacks) and direct sensor attacks (e.g., sensor spoofing).} 
\label{fig:architecture}
\end{figure}

In this section, we formalize the problem considered in this work.  We start from 
the security-aware system model (i.e., including the attack impact)  illustrated in Fig.~\ref{fig:architecture}, with each component  described in detail as follows. 

\subsection{System Model}
\label{sec:model}

We consider general nonlinear 
dynamics of a physical system (i.e., plant) compromised by attacks on system sensing, modeled~as
%
\begin{equation}
\label{eq:NLplant}
\begin{split}
x_{t+1} &=f(x_{t},u_{t})+w_{t}, \\ 
y_t^c &=y_t+a_t=h(x_t)+v_t+a_t.
\end{split}
\end{equation}

Here, $x_t\in \mathbb{R}^n$ and $u_t\in \mathbb{R}^m$ denote the plant’s state and input vectors at time $t$, whereas the  output vector received by the controller $y_t^c\in \mathbb{R}^p$ 
contains the measurements 
from $p$ sensors from the set $\mathcal{S}=\{s_1,...,s_p\}$, including compromised measurements provided by sensors from the set $\mathcal{K}_a\subseteq\mathcal{S}$; $a_t\in\mathbb{R}^p$ denotes the attack signal injected at time $t$, and thus the vector is sparse with support in $\mathcal{K}_a$ -- i.e., $\text{supp}(a_t)=\mathcal{K}_a$ and $a_{t,i}=0$ for $i\in\mathcal{K}_a^\complement$.\footnote{We refer to sensors from $\mathcal{K}_a$ as compromised, even if a sensor itself is not directly compromised but its measurements may be altered due to e.g., network-based attacks. 
}
%
%
The observation function $h:\mathbb{R}^{n}\to \mathbb{R}^p$ is assumed to be $L$-Lipschitz.  
Finally, $w_t$ and $v_t$ are the state and measurement noise, respectfully. 
%
%
%
%

In a special case, if the plant~\eqref{eq:NLplant} is linear time-invariant (LTI), we use $f(x_t,u_t)=Ax_t+Bu_t$ and $h(x_t)=Cx_t$, 
where $A$, $B$ and $C$ are matrices of suitable dimensions.

\vspace{4pt}
\noindent\textbf{Control Architecture.} 
We consider a common control architecture, 
with 
three main components (as illustrated in Fig.~\ref{fig:architecture}):~a~state estimator, a feedback controller, and an anomaly detector. 

\vspace{2pt}
The \textbf{\emph{State Estimator}}
(observer) employs the system model to predict its (state) evolution, and thus provide an estimate $\hat{x}_t$ of its state at time $t$; in general, this can be captured as
\begin{equation}\label{eq:estimator}
\begin{split}
\hat{x}_{t} &= \mathcal{}{O}_t(\hat{x}_{t-1},u_{t-1},y_t), \quad \hat{y}_t = h(\hat{x}_t).
\end{split}
\end{equation}

The mapping $O_t$ is commonly designed so that~\eqref{eq:estimator} minimizes a norm of the estimation~error defined~as 
\begin{equation}
\label{eq:Delta}
\Delta{x}_t=x_t-\hat{x}_t.
\end{equation}

Depending on the system model and statistical characteristics of the noise, different estimation methods may be employed. 
Kalman filters are widely used for LTI systems, 
whereas Extended Kalman filters (EKFs) are mainly utilized for nonlinear systems with Gaussian noise, e.g., the autonomous driving and 
UAV applications considered in this work. 
%
Thus, we particularly focus~on~EKFs. 
The EKF functionality 
for a system~\eqref{eq:NLplant} is described~by
\begin{align}
\hat{x}_{t|t-1} = f(\hat{x}_{t-1},u_{t-1}), ~
\hat{x}_t = \hat{x}_{t|t-1}+L_t(y_t-h(\hat{x}_{t|t-1})),\nonumber \hat{y}_t = h(\hat{x}_t);
\end{align}
here, $\hat{x}_{t|t-1}$, $\hat{x}_{t}$ and $\hat{y}_t$ denote the predicted state estimate, (updated) state estimate, and predicted output, respectively. The EKF gain $L_t$ is also updated as 
\begin{equation}\label{eq:Ricatti}
\begin{split}
L_t &= A_tP_tC_t^T\big(C_tP_tC_t^T+R\big)^{-1},\\
P_{t+1} &= A_tP_tA_t^T+Q-L_t\big(C_tP_tC_t^T+R\big)L_t^T,
\end{split}
\end{equation}
where $A_t=\frac{\partial f(x_t,u_t)}{\partial x_t}|_{\hat{x}_{t-1},u_t}$ and $C_t=\frac{\partial h(x_t)}{\partial x_t}|_{\hat{x}_{t|t-1}}$ are the Taylor expansion of $f$ and $h$ around $(\hat{x}_{t-1},u_t)$ and $\hat{x}_{t|t-1}$, respectively.~Also, $Q$ and $R$ are the covariance matrices of the Gaussian 
noises $w_t$ and $v_t$, respectively. 
Finally, the residue signal (or innovation noise) is defined~as
\begin{equation}
\label{eq:residue}
z_t=y_t-h(\hat{x}_{t|t-1}).
\end{equation}
For systems with Gaussian noise, its covariance matrix 
is 
$S_t=\mathbb{E}\{z_tz_t^T\}=C_tP_tC_t^T+R_t$~\cite{julier2004unscented}.

\vspace{4pt}
The \emph{\textbf{Feedback Controller}} employs the control law
$u_t=\pi(\hat{x}_t)$; 
without loss of generality, we assume the control goal is to regulate the states to $0\in\mathbb{R}^n$. 
Hence, the estimator~\eqref{eq:estimator} can be modeled~as
\begin{equation}
\label{eq:estimK}
\hat{x}_{t} = {O}_t(\hat{x}_{t-1},\pi(\hat{x}_{t-1}),y_t)\delequal\mathcal{O}_t(\hat{x}_{t-1},y_t).
\end{equation}

\vspace{2pt}
The \emph{\textbf{Anomaly Detector}} (AD) is used to detect the presence of system anomalies, including intrusions (i.e., attacks). 
The standard approach 
is to use the system model to predict the future system behavior and compare it with the actual observation (e.g., see~\cite{giraldo2018survey} and the references within); capturing the discrepancy between the system and its predicted behavior with a  detection function~$g_t$. 

In feedback-control based CPS, the residue~\eqref{eq:residue} is widely used for anomaly detection --  $\chi^2$ detector in~\cite{mo2010false,mo2015performance}, cumulative sum in~\cite{tunga2018tuning}, sequential probability ratio test (SPRT) detector in~\cite{kwon2016real}, and a general window-type detector in~\cite{jovanov_tac19}. 
For instance, for $\chi^2$-based detectors, 
the detection function $g_t$ is a weighted norm of $z_t$ (with the $\chi^2$ distribution) -- i.e., 
\begin{equation}
\label{eq:gt}
g_t=z_t^TS_t^{-1}z_t;
\end{equation}
the other detectors (e.g.,~from~\cite{kwon2014analysis,kwon2016real,mo2015performance,kwon2017reachability,tunga2018tuning,mo2010false,miao_cdc13,miao_tcns17,jovanov_tac19}) 
use some forms of a windowed extension of~\eqref{eq:gt}.  
Therefore, to simplify our presentation, we focus on 
the detection function $g_t$ from~\eqref{eq:gt}, and our results can be directly extended to other cases.

Finally, the system triggers alarm if the detection function satisfies that 
$g_t>\eta,$
for some predefined threshold value~$\eta$.
Usually the value $\eta$ is assigned such that under normal conditions (i.e., when the system is not compromised) it holds that $\mathbb{P}(g_t >\eta)\leq\epsilon$ -- i.e., the system has a low false alarm rate. 

\subsection{Attack Model}

We assume that the attacker has access to the system (or an instance or the model of the system) offline, allowing offline design of a suitable attack generator, which is then employed at runtime to degrade system operation by compromising the sensing~measurements. 

\paragraph*{Attacker capabilities during offline training.}
Let $T$ be the duration of the training phase; we define $\hat{\mathbf{ X}}_{t|t-1}=\{\hat{x}_{0|-1} , \hat{x}_{1|0},...,\hat{x}_{t|t-1} \} $,  $\mathbf{Y}_t=\{y_{0},...,y_t \}$ and $\mathbf{L}_t=\{L_{0},...,L_{t}\}$ as the sequences of the predicted states,  plant outputs, and EKF gains for $t\geq 0$, with $t=0$ denoting the time starting the training phase. We assume that the attacker has access to the EKF values over time (either directly, or knowing the EKF design and running a copy of the EKF in parallel) -- i.e., 
has access to $\hat{\mathbf{ X}}_{T|T-1}$, $\mathbf{L}_T$, $\mathbf{Y}_T$ and the function $h$; {specifically, the attacker does not need to know the actual function $h$, but rather its potential approximation used in~\eqref{eq:estimator} to implement the state estimator.}
Meanwhile, for $0\leq t<T$, the attacker can compromise the sensor measurements according to the model from~\eqref{eq:NLplant}.\footnote{In general, the training (i.e., offline) time is different than the run-time $t$, as offline training and runtime-deployment are performed on different instances of the system~\eqref{eq:NLplant}. However, to simplify our notation we do not differentiate between these time axes, unless the use of specific time (offline vs. runtime) is not clear from the context.}


\paragraph*{Attacker capabilities at runtime -- i.e., during attack.}
Let $t_0$ denote the start time of the attack, modeled as in~\eqref{eq:NLplant}, and $T'$ its duration. 
Again, we assume that the attacker has access to the 
sensor measurements $y_t$. 
In addition, we will consider two attack scenarios: when the attacker (\textit{i})~\textbf{does (i.e., grey-box)}, or (\textit{ii})~\textbf{does not (i.e., black-box)} have access to 
the state estimation $\hat{x}_{t-1}$ in the previous time step; the latter threat model is especially impactful, as it assumes that \textit{the attacker does not have access to the internal controller variables at runtime, but only measurements from the 
(compromised) system~sensors.} 

\paragraph{{Attacker's goal}}is to 
%
\emph{maximize degradation of control performance -- i.e., QoC}. Specifically, as only sensor data may be compromised, the attack objective is to \emph{maximize the state estimation error}~$\Delta x_t$. 
In addition, the attacker wants \emph{to remain stealthy -- i.e., undetected by the anomaly detector}. 
These notions are formalized as~follows.



\begin{definition}
\label{def:eps_alpha}
The sequence of attack vectors $a_{t_0},a_{t_0+1},...$ is referred to as $(\epsilon,\alpha)$-successful if there exists $t_0 \leq t'\leq T'+t_0$ such that $ \Vert \Delta{x}_{t'} \Vert \geq \alpha$  and $\mathbb{P}( g_t >\eta) \leq \epsilon$ for all $t_0 \leq t\leq T'+t_0$.\footnote{Note that any $p$-norm can be used. In this work, when $p$ is not specified, the use of the standard 2-norm is implied.} 
\end{definition}

Therefore, the attacker's goal is to insert a sequence of false data measurements $a_{t_0},...,a_{t_0+T'}$ resulting in an $(\epsilon,\alpha)$-successful attack. 
Note that while~\cref{def:eps_alpha} focuses on attacks that result in a desired norm of the estimation error (i.e., greater than $\alpha$), for some systems, attacks may cause arbitrarily large estimation errors~\cite{jovanov_tac19,mo2010false,kwon2014analysis}. 
%
%
For LTI systems with (standard) Kalman filters,
the notion of $(\epsilon,\alpha)$-successful attacks was first introduced in~\cite{mo2010false}. Also, for LTI systems necessary and sufficient conditions such that $(\epsilon,\alpha)$-successful attacks exist for any $\alpha >0$ are introduced in~\cite{mo2010false,kwon2014analysis,jovanov_tac19}, along with methods to derive such attacks. However, to the best of our knowledge, there is no 
method for vulnerability analysis  of nonlinear systems from~\eqref{eq:NLplant} under sensor-based~attacks;~i.e.,~the~impact that such attacks would have on the estimation error, and~thus~QoC. 

\section{Adversarial Learning for Nonlinear Dynamical Systems} 
\label{sec:stealthy} 
In this section, we present methods to design attack-generators for stealthy and effective sensing attacks. Before considering general nonlinear dynamics, we motivate the considered approaches by considering attacks on LTI systems; 
we start with design 
of $(\epsilon,\alpha)$-successful attacks against LTI systems with standard Kalman~filters. 

\begin{lemma}[{\cite{mo2010false,kwon2014analysis,jovanov_tac19}}]
\label{lem:PA1}
For an LTI system 
with a Kalman filter-based estimator, there exist $(\epsilon,\alpha)$-successful attacks for any desired $\alpha>0$ 
if and only if the matrix $A$ is unstable and at least one eigenvector $v$ corresponding to an unstable eigenvalue satisfies that $\text{supp}(Cv)\subseteq \mathcal{K}_a$. 
\end{lemma}

Note that $A$ being unstable is a necessary condition for existence of $(\epsilon,\alpha)$-successful attack for arbitrarily large $\alpha$, in 
LTI systems with Kalman filters. 
However, if all sensors are under attack (i.e., $\mathcal{K}_a=\mathcal{S}$), this is also a sufficient condition.  A similar necessary and sufficient condition \textbf{only} for LTI systems with bounded noise that employ novel attack-resilient estimators (e.g., from~\cite{fawzi2014secure,pajic_csm17,pajic_tcns17}) is derived in~\cite{khazraei_automatica21}. 
%
%
Now, we show how to use only the current state estimation $\hat{x}_t$, the plant output $y_t$ and input $u_{t-1}$ to compute such attack sequence on LTI systems for arbitrarily large $\alpha$. 

\begin{theorem}\label{Thm:PA}
Consider an LTI system  with unstable matrix $A$, and let $\phi_t$ denote a Gaussian noise vector satisfying $\mathbb{E}\{\phi_t\}=0$ and $\mathbb{E}\{\phi_t\phi_t^T\}\preceq S$. 
The attack sequence generated by $a_t=-y_t+CBu_{t-1}+CA\hat{x}_{t-1}+\phi_t$, for $t\geq 0$, is an $(\epsilon,\alpha)$-successful attack for any 
$\alpha>0$.
\end{theorem}

Proof of the theorem is available in Appendix~\ref{app:T1proof}.
%

%
%
%
%
Unlike in design of adversarial examples for images      (e.g.,~\cite{szegedy2013intriguing,goodfellow2014explaining,papernot2017practical,yuan2019adversarial}), Theorem~\ref{Thm:PA} shows that for LTI dynamical systems, an effective 
and stealthy attack sequence has to evolve over time (i.e., following suitable dynamics). 
%
This is well aligned with stealthy attack design methods from~\cite{mo2010false,kwon2014analysis} that do not directly analyze the attack impact on the LTI system but rather consider a dynamical system capturing the difference between the evolutions of the non-attacked and compromised systems.

The fact that effective stealthy attacks should follow certain dynamics (i.e., evolve over time)
inspires us to use a similar structure for generating effective stealthy attacks against system with nonlinear dynamics. 
%
Specifically, 
we consider attacks with dynamics $a_t=F(\hat{x}_{t-1},y_t)$ (note that  $y_t$ and $\hat{x}_{t-1}$ are both  functions of~$a_{t-1}$), with the idea to use 
a DNN to learn $F$.
%
%
However, the attacker might not always have access to $\hat{x}_{t-1}$ during the attack; {for example, if (s)he used an instance 
of the system to train the generator, and then uses the derived attack signals to insert attacks over the network without access to the internal execution context of the new system (instance) under attack.}  

Thus, we will consider both cases where information $\hat{x}_{t-1}$ is (\textit{i})~available to the attacker at runtime, and (\textit{ii})~when it is not. For the latter, the idea is to \textit{replace $\hat{x}_t$ with another signal $r_t$ that is directly constructed by the attacker}. We now summarize 
our models to design $(\epsilon,\alpha)$-successful attack generators for nonlinear systems.

\paragraph{\textbf{FNN-based attack design}.} When $\hat{x}_{t-1}$ is available to the attacker, we design an FNN that uses the current output measurements and the last state estimation to generate the next~attack~vector (for current time) 
-- i.e., 
%
\begin{equation}\label{eq:FNN}
a_t=H_{\theta}(y_t,\hat{x}_{t-1}),
\end{equation}
where $H_{\theta}$ is a 
deep FNN with parameters $\theta$, input dimension of $n+p$ and output dimension $p$.

\paragraph{\textbf{RNN-based attack design.}}
When $\hat{x}_{t-1}$ is not available to the attacker, we consider an RNN architecture, that uses only the 
current sensor measurements for attack design as follows
\begin{equation}\label{eq:RNN}
\begin{split}
r_t &= G_{\theta}(y_{t},r_{t-1}), \\
a_t &= W r_t;
\end{split}
\end{equation}
here, $G_{\theta}$  is the state update function  
implemented as an RNN
parameterized by $\theta$, while $W\in \mathbb{R}^{l\times p}$ is a linear mapping from the state $r_t$ to the attack vector $a_t$. Intuitively, $r_t\in \mathbb{R}^{l}$ and $G_{\theta}$ should allow for 
capturing of the evolution of $\hat{x}_{t}$, 
to create a dynamical pattern for the sequence of attack vectors in the RNN design~\eqref{eq:RNN}. 

\paragraph{\textbf{Model training.}} To capture the attack impact on the estimation error $\Delta x_t$ from~\eqref{eq:Delta}, the first challenge 
is that the actual true system state  $x_t$ is not available to the attacker; instead, only sensor measurements $y_t$ are known. We show that under some mild assumptions, the sensor measurements can be directly~used. 

\begin{theorem}
\label{thm:bound}
Consider the system~\eqref{eq:NLplant}.
If the function $h:\mathbb{R}^n\to \mathbb{R}^p$ is Lipschitz with constant $L$, then $ \Vert y_t-h(\hat{x}_t) \Vert \geq \alpha $ implies $ \Vert x_t-\hat{x}_t \Vert \geq \frac{\alpha - \sqrt{\sigma}k}{L} $  with probability $1-\frac{p}{k^2}$, for any $k$ such that $k<\frac{\alpha}{\sigma}$ and $R\preceq \sigma I$; here, $R$ is the covariance matrices of the Gaussian measurement noise $v_t$,
$I$ is the identity matrix, and $\sigma$ is a positive scalar.
\end{theorem}

Proof of \cref{thm:bound} is provided in Appendix~\ref{app:T2proof}.

We use $a_t=F_{\Theta}(y_t,s_{t-1})$ to capture the attack vectors generated by either \eqref{eq:FNN} or \eqref{eq:RNN}; here,  $s_t$ indicates either $r_t$ or $\hat{x}_t$. 
Our goal is to train the parameters 
in \eqref{eq:FNN} and \eqref{eq:RNN} so that 
the networks act as generators of $(\epsilon,\alpha)$-successful attacks. 
To achieve this, we use the following~approach. 

Starting offline  training at time $t=0$, we seek for $a_{0}$ that maximizes $\mathbb{E} \Vert x_{0}-\hat{x}_{0}\Vert ^2$ with $\hat{x}_{0}=\mathcal{O}(\hat{x}_{-1},y_{0}^c)$ and $y^c_{0}=y_{0}+a_{0}$; here, the expected value operation $\mathbb{E}\{\cdot \}$ is over random variables $w$ and $v$. 
As the true state $x_{0}$ is not available, from Theorem~\ref{thm:bound}, 
we can maximize 
$\mathbb{E} \Vert y_{0}-h(\hat{x}_{0}) \Vert ^2 $ instead. 
Also, the generated attack should satisfy the stealthiness condition $\mathbb{P}(  g_{0 } >\eta) \leq \epsilon$, with $g_{0}$ defined in~\eqref{eq:gt} for $z_0=y^c_{0}-\hat{y}_{0}$;  for $\chi^2$-based ADs, it holds that $\hat{y}_{0}=h(\hat{x}_{{0}|-1})$. Thus, the following optimization problem should be solved 
at time $t=0$
%
\begin{equation*}
\begin{split}
&\max_{\Theta}\mathbb{E} \Vert  y_{0}-h(\hat{x}_{0}) \Vert ^2\\
 & \,\, s.t. \,\, \mathbb{P}(  g_{0 } >\eta) \leq \epsilon\\
 & \,\, a_{0}=F_{\Theta}(y_{0},s_{-1}).
\end{split}
\end{equation*}
As this constrained optimization problem is challenging to solve, we penalize the norm of the residue signal and incorporate a new term in our objective function, 
resulting in the optimization problem:
%
\begin{equation}\label{eq:objective1}
\begin{split}
&\min_{\Theta} \mathbb{E}\{g_{0} -\delta \Vert y_{0}-h(\hat{x}_{0}) \Vert ^2\}\\
 & \,\, a_{0}=F_{\Theta}(y_{0},s_{-1});
\end{split}
\end{equation}
%
%
here, $\delta>0$ is a standard regularization term balancing the stealthiness condition and performance degradation caused by the estimation error. 

Let us denote the parameters obtained 
from~\eqref{eq:objective1} as $\Theta^{(0)}$. Now, the attack vector $a_{0}=F_{\Theta^{(0)}}(y_{0},s_{0})$  applied to the sensor measurements at time $t=0$, 
would results in state estimate $\hat{x}_{0}=\mathcal{O}(\hat{x}_{-1},y_{0}^c)$. 
In the next (offline) time step ($t=1$), we search for parameters 
such that $ \mathbb{E} \{g_{1} -\delta \Vert y_{1}-h(\hat{x}_{1}) \Vert ^2 \}$ is minimized. However, in this case the parameters will only be trained to minimize the cost function at time $t=1$ and thus will disregard minimization of the cost function in the previous time step. Hence, the cost function from the previous time step should be also included -- i.e., the objective function to be minimized at time $t=1$ should be be $$\mathbb{E} \{g_{0} +g_{t_0+1} -\delta \left(\Vert y_{0}-h(\hat{x}_{0}) \Vert ^2 + \Vert y_{1}-h(\hat{x}_{1}) \Vert ^2 \right)\}.$$ 

This approach should continue for the following (offline) time steps. Generally, if we consider that the training starts at $t=0$, for any  $t\geq 0$ there should be an instantaneous cost function defined by $J_t=g_{t}  -\delta \Vert y_{t}-h(\hat{x}_{t}) \Vert$. 
Therefore, the offline optimization problem that is solved at time step $t$ is 
\begin{equation}\label{eq:objective2}
\begin{split}
&\min_{\Theta} \mathbb{E} \{J_t +\lambda_t \sum\nolimits_{j=0}^{t-1} J_j\} \\
 & \,\, a_{t}=F_{\Theta}(y_{t},s_{t-1}).
\end{split}
\end{equation}
Again, $\lambda_t\geq 0$ are regularization terms to control the incorporation of previous 
cost functions. If $\lambda_t=1$, we effectively penalize all previous and current instantaneous costs equally. For smaller values of $\lambda_t$, the cost function at time $t$ will be approximately $J_t$; 
i.e., we can do more exploration by only minimizing the cost at time $t$. However, increasing $\lambda_t$ helps exploit more by giving more importance to the previous cost functions. 

\paragraph{\textbf{Training Algorithm}.} In our proposed algorithms, we use Monte Carlo (MC) averaging to approximate the expectation in the cost function~\eqref{eq:objective2} by the sample mean over $N$ number of simulated trajectories starting at time $t=0$. Specifically, the $i$-th,  {$i=1,...,N$,} trajectory at time $t$ is obtained~by
\begin{equation*}
x_{t}^{i}=f(x_{t-1}^{i},u_{t-1}^{i})+w_{t-1}^{i},\quad y_t^{i}=h(x_{t}^{i})+v_t^{i},\quad \hat{x}_{t|t-1}^{i}=f(\hat{x}_{t-1}^{i},u_{t-1}^{i}).
\end{equation*}

The cost that the trajectory $i$ at time $t$ imposes is $J_t^i=g_{t}^i  -\delta \Vert y_{t}^i-h(\hat{x}_{t}^i) \Vert$. However, at time step $t$ the DNN is trained such that $J'_t=\frac{1}{N}\sum_{i=1}^{N} \big(\lambda_t\sum_{j=0}^{t-1}  J_j^i +  J_t^i\big)$ is minimized, where $J'_t$ is the approximated expectation of cost function in~\eqref{eq:objective2}. Finally, once the model parameters 
are obtained at time $t$, the attack vector $a_t^i=F_{\Theta^{(t)}}(y_t^i,s_{t-1}^i)$ is applied to the system output $y_t^i$, 
and the process is repeated until the training completes.
%

When the 
EKF is used,  Algorithm~\ref{alg1} captures pseudocode for learning the FNN-based stealthy attack generator, and Algorithm~\ref{alg:the_alg2} 
summarizes pseudocode to learn the RNN-based attack generators. 


\begin{algorithm}[!t]
  \caption{Learning Stealthy and Effective FNN-based Attack-Generator Models, using MC Simulation}
  \label{alg1}
  \begin{algorithmic}[1] 
  \State {Set the learning rate $\beta$, training period $T$, sample number $N$}
    \For{$t=0:T$}
        \State $x_{t}^{1:N}=f(x_{t-1}^{1:N},u_{t-1}^{1:N})+w_{t-1}^{1:N}$,~~~
        $y_t^{1:N}=h(x_{t}^{1:N})+v_t^{1:N}$
        \State $\hat{x}_{t|t-1}^{1:N}=f(\hat{x}_{t-1}^{1:N},u_{t-1}^{1:N})$
        \Repeat
        \State $J'_t=\frac{1}{N} \sum_{i=1}^{N} \big(\lambda_t\sum_{j=0}^{t-1}  J_j^i +  J_t^i\big) $ with ${y_j^c}^{1:N}=y_j^{1:N}+H_{\theta}(\hat{x}_{j-1}^{1:N},{y_j^{1:N}})$
        \State $\theta^{(t)}\leftarrow \theta^{(t)}-\beta \nabla_{\theta} J'_t$
        \Until{{Convergence}}
        \State $a_t^{1:N}=H_{\theta^{(t)}}(\hat{x}_{j-1}^{1:N},{y_j^{1:N}})$,~~~
        ${y_t^c}^{1:N}=y_t^{1:N}+a_t^{1:N}$
        \State $\hat{x}_t^{1:N}=\hat{x}_{t|t-1}^{1:N}+L_t^{1:N}({y_j^c}^{1:N}-h(\hat{x}_{t|t-1}^{1:N}))$,~~~
        $u_t^{1:N}=\pi(\hat{x}_t^{1:N})$
    \EndFor
  \end{algorithmic}
\end{algorithm}

\begin{algorithm}[!t]
  \caption{Learning Stealthy and Effective RNN-based Attack Generator Models, using MC Simulation}
  \label{alg:the_alg2}
  \begin{algorithmic}[1]
  \State {Set the learning rate $\beta$, training period $T$, sample number $N$}
    \For{$t=0:T$}
        \State $x_t^{1:N}=f(x_{t-1}^{1:N},u_{t-1}^{1:N})+w_{t-1}^{1:N}$
        \State $y_t^{1:N}=h(x_{t}^{1:N})+v_t^{1:N}$
        \State $\hat{x}_{t|t-1}^{1:N}=f(\hat{x}_{t-1}^{1:N},u_{t-1}^{1:N})$
        \Repeat
        \State $J'_t=\frac{1}{N} \sum_{i=1}^{N} \big(\lambda_t\sum_{j=0}^{t-1}  J_j^i +  J_t^i\big) $ with ${y_t^c}^{1:N}=y_j^{1:N}+WG_{\theta}(r_{j-1}^{1:N},{y_j}^{1:N})$
        \State $\theta^{(t)}\leftarrow \theta^{(t)}-\beta \nabla_{\theta} J'_t$
        \State $W^{(t)}\leftarrow W^{(t)}-\beta \nabla_{W} J'_t$
        \Until{Convergence}
        \State $r_t^{1:N}=G_{\theta^{(t)}}(r_{t-1}^{1:N},{y_t^{1:N}})$
        \State $a_t^{1:N}=W^{(t)}r_t^{1:N}$
        \State ${y_t^c}^{1:N}=y_t^{1:N}+a_t^{1:N}$
        \State $\hat{x}_t^{1:N}=\hat{x}_{t|t-1}^{1:N}+L_t^{1:N}({y_t^c}^{1:N}-h(\hat{x}_{t|t-1}^{1:N}))$
        \State $u_t^{1:N}=\pi(\hat{x}_t^{1:N})$
      \EndFor
  \end{algorithmic}
\end{algorithm}


\section{Case Studies}  \label{sec:exp} 

We illustrate and thoroughly evaluate our attack-design framework on 
three case studies, inverted pendulum, autonomous driving vehicles (ADVs) and unmanned aerial vehicles (UAVs), with varying level of complexity due to system dynamics. 
For all case studies, the workstation used for training is powered by Nvidia RTX 6000 GPUs with 24~GB of memory each, two Intel Xeon Silver 4208 CPUs with 16 cores each, and a total of 192 GB RAM. The code was developed using Python and Pytorch deep learning libraries.

\subsection{Inverted Pendulum}

To illustrate the effectiveness of our algorithm compared to the LTI-based methods from \cite{mo2010false,jovanov_tac19}, we first considered a fixed-base inverted pendulum. We used the nonlinear dynamical model 
from~\cite{formal2006inverted}. Two sensors were used to measure the states of the system, $\theta$ and $\dot{\theta}$, where $\theta$ was the angle of pendulum rod from the vertical axis measured counterclockwise. The threshold $\eta$ was set to have $\epsilon=.01$ (i.e., 
on average, every one hundred time steps a false alarm occurs). A feedback controller was used to keep the pendulum inverted around $\theta=0$ equilibrium~point. 

\begin{table}[!t]
    \centering
    \caption{Attack success rate (SR) for different training duration $T$ with $N=200$, $\lambda=.5$ and varying $\alpha$ and $\theta$ over 100 inverted pendulum experiments; FNN/RNN  denote FNN vs. RNN success~rates.}
    \vspace{-4pt}
    \setlength\tabcolsep{0.pt}
    \begin{tabular}{c|c|c|c|c|c|c|c|c|c|c}
    \hline
    \makecell{\footnotesize	 SR \% \\ \footnotesize	 FNN/RNN} & \footnotesize	 $\alpha=.5$ &  \footnotesize	 $\alpha=1$  & \footnotesize	  $\alpha=2$ & \footnotesize	  $\alpha=4$ & \footnotesize	  $\alpha=8$ & \footnotesize	  $|\theta|=\frac{\pi}{8}$ & \footnotesize	  $|\theta|=\frac{\pi}{4}$ & \footnotesize	  $|\theta|=\frac{\pi}{3}$ & \footnotesize	  $|\theta|=\frac{\pi}{2}$ & \footnotesize	  $|\theta|=\pi$
     \\
    \hline
     \footnotesize	 $T=50$    & \footnotesize	 0/0 & \footnotesize	 {0/0}	& \footnotesize	 {0/0} & \footnotesize	 0/0	& \footnotesize	 {0/0} & \footnotesize	0/0 & \footnotesize	 0/0 & \footnotesize	 0/0 & \footnotesize	 0/0 & \footnotesize	 0/0\\
      \footnotesize	 $T=80$    & \footnotesize	{100/99} & \footnotesize	 26/67	& \footnotesize	{0/0} & \footnotesize	 0/0	& \footnotesize	{0/0} & \footnotesize	 0/0 & \footnotesize	 0/0 & \footnotesize	 0/0 & \footnotesize	 0/0 & \footnotesize	 0/0\\
     \footnotesize	 $T=100$    & \footnotesize	100/100 & \footnotesize	{100/100}	& \footnotesize	100/78 & \footnotesize	{0/3}	& \footnotesize	{0/0} & \footnotesize	 100/81 & \footnotesize	 0/9 & \footnotesize	 0/0 & \footnotesize	 0/0 & \footnotesize	 0/0\\
     \footnotesize	 $T=120$    & \footnotesize	 100/100 & \footnotesize	{100/100} & \footnotesize	{100/100} & \footnotesize	100/100& \footnotesize	{0/0} & \footnotesize	100/100 & \footnotesize	 100/100 & \footnotesize	 100/99 & \footnotesize	 0/52 & \footnotesize	 0/0\\
     \footnotesize	 $T=150$    & \footnotesize	 100/100 & \footnotesize	{100/100} & \footnotesize	{100/100} & \footnotesize	100/100& \footnotesize	{89/97} & \footnotesize	100/100 & \footnotesize	 100/100 & \footnotesize	 100/100 & \footnotesize	 100/100 & \footnotesize	 50/58\\
     \footnotesize	 $T=170$    & \footnotesize	 100/100 & \footnotesize	{100/100} & \footnotesize	{100/100} & \footnotesize	100/100& \footnotesize	{100/100} & \footnotesize	100/100 & \footnotesize	 100/100 & \footnotesize	 100/100 & \footnotesize	 100/100 & \footnotesize	 100/100\\
    \hline 
    \makecell{ \footnotesize	 SR \% \\\footnotesize LTI Model} & \footnotesize	 100 & \footnotesize	 100 & \footnotesize	 100 & \footnotesize	 10 & \footnotesize	 0 & \footnotesize	 100 & \footnotesize	 14 & \footnotesize	 0 & \footnotesize	 0 & \footnotesize	 0
     \\
    \hline\hline
    \end{tabular}
    \label{tab:duration}
\end{table}

\begin{table}[!t]
    \centering
    \caption{Attack success rate (SR) for different values of $N$ with $T=150$, $\lambda=.5$ and different values of $\alpha$ and $\theta$ over 100 experiments for the inverted pendulum. The FNN/RNN numbers denote FNN vs. RNN success rate values.}
    \vspace{-4pt}
    \setlength\tabcolsep{0.pt}
    \begin{tabular}{c|c|c|c|c|c|c|c|c|c|c}
    \hline
    \makecell{\footnotesize	 SR \% \\ \footnotesize	 FNN/RNN} & \footnotesize	 $\alpha=.5$ &\footnotesize	 $\alpha=1$  & \footnotesize	  $\alpha=2$ & \footnotesize	  $\alpha=4$ & \footnotesize	  $\alpha=8$ & \footnotesize	  $|\theta|=\frac{\pi}{8}$ & \footnotesize	  $|\theta|=\frac{\pi}{4}$ & \footnotesize	  $|\theta|=\frac{\pi}{3}$ & \footnotesize	  $|\theta|=\frac{\pi}{2}$ & \footnotesize	  $|\theta|=\pi$
     \\
    \hline
     \footnotesize	 $N=1$    & \footnotesize	 35/4 & \footnotesize	 {35/2}	& \footnotesize	 {35/1} & \footnotesize	 35/1	& \footnotesize	 {10/0} & \footnotesize	 35/1 & \footnotesize	 35/1 & \footnotesize	 35/1 & \footnotesize	 35/0 & \footnotesize	 0/0\\
      \footnotesize	 $N=10$    & \footnotesize	{100/19} & \footnotesize	 100/10	& \footnotesize	{100/4} & \footnotesize	 100/2	& \footnotesize	{45/1} & \footnotesize	 100/4 & \footnotesize	 100/2 & \footnotesize	 100/1 & \footnotesize	 100/1 & \footnotesize	 0/0\\
     \footnotesize	 $N=50$    & \footnotesize	100/74 & \footnotesize	{100/66}	& \footnotesize	100/56 & \footnotesize	{100/38}	& \footnotesize	{100/4} & \footnotesize	 100/56 & \footnotesize	 100/39 & \footnotesize	 100/32 & \footnotesize	 100/9 & \footnotesize	 0/0\\
     \footnotesize	 $N=100$    & \footnotesize	 100/96 & \footnotesize	{100/96} & \footnotesize	{100/96} & \footnotesize	 100/95& \footnotesize	{90/64} & \footnotesize	 100/96 & \footnotesize	 100/96 & \footnotesize	 100/95 & \footnotesize	 100/90 & \footnotesize	 15/1\\
     \footnotesize	 $N=250$    & \footnotesize	 100/100 & \footnotesize	{100/100} & \footnotesize	{100/100} & \footnotesize	 100/100& \footnotesize	{95/97} & \footnotesize	 100/100 & \footnotesize	 100/100 & \footnotesize	 100/100 & \footnotesize	 100/100 & \footnotesize	 94/95\\
     \footnotesize	 $N=500$    & \footnotesize	 100/100 & \footnotesize	{100/100} & \footnotesize	{100/100} & \footnotesize	 100/100& \footnotesize	{100/100} & \footnotesize	 100/100 & \footnotesize	 100/100 & \footnotesize	 100/100 & \footnotesize	 100/100 & \footnotesize	 100/99\\
    \hline\hline
    \end{tabular}
    \label{tab:MC_sample}
\end{table}



\begin{table}[!t]
    \centering
    \caption{Attack success rate (SR) for different values of $\lambda$ with $T=100$, $N=200$ and different values of $\alpha$ and $\theta$ over 100 experiments for the inverted pendulum. The FNN/RNN numbers denote FNN vs RNN success rate values.}
    \vspace{-4pt}
    \setlength\tabcolsep{0.pt}
    \begin{tabular}{c|c|c|c|c|c|c|c|c|c}
    \hline
    \makecell{\small SR \% \\ \small FNN/RNN} & \small $\alpha=.5$ &\small $\alpha=1$  & \small  $\alpha=2$ & \small  $\alpha=4$ & \small  $\alpha=8$ & \small  $|\theta|=\frac{\pi}{8}$ & \small  $|\theta|=\frac{\pi}{4}$ & \small  $|\theta|=\frac{\pi}{3}$ & \small  $|\theta|=\frac{\pi}{2}$
     \\
    \hline
     \small $\lambda=.01$    & \small 100/12 & \small {100/2}	& \small {100/1} & \small 50/1	& \small {0/0} & \small 100/1 & \small 50/1 & \small 0/0 & \small 0/0 \\
      \small $\lambda=.05$    & \small{100/98} & \small 100/97	& \small{100/90} & \small 58/0	& \small{0/0} & \small 100/92 & \small 53/0 & \small 0/0 & \small 0/0 \\
     \small $\lambda=.1$    & \small100/98 & \small{100/97}	& \small100/91 & \small{73/38}	& \small{0/0} & \small 100/93 & \small 55/45 & \small 0/8 & \small 0/0  \\
     \small $\lambda=1$    & \small 100/100 & \small{100/100} & \small{100/70} & \small 97/59 & \small{0/0} & \small 100/75 & \small 100/60 & \small 54/3 & \small 0/0 \\
     \small $\lambda=5$    & \small 100/100 & \small{100/100} & \small{100/100} & \small 70/67& \small{0/0} & \small 100/100 & \small 90/78 & \small 10/27 & \small 0/0 \\
      \small $\lambda=100$    & \small 100/100 & \small{100/64} & \small{100/42} & \small 0/12& \small{0/0} & \small 100/43 & \small 1/16 & \small 0/0 & \small 0/0 \\
    \hline\hline
    \end{tabular}
    \label{tab:lambda}
\end{table}

First, we trained both FNN and RNN models with $N=200$ MC samples, $\lambda=.5$ and different values of training period $T$ {when both sensors are compromises}. Table 
\ref{tab:duration} shows the success rate (in percentage) of both attack generator models, which were obtained for each training period $T$, and for different values of $\alpha$ and $\theta$. For example, consider the entry associated with row $T=150$ and $\alpha=8$. It shows that 89 percent of times the FNN and 97 percent of times the RNN model was successful in driving the system to $\alpha=8$, with $\alpha$ being the norm of the state estimation error. This 
also applies to the columns shown by $|\theta|$; for example, the entry in row $T=150$ and column $|\theta|=\frac{\pi}{2}$ means that 100 percent of times both FNN and RNN models were successful in driving the pendulum rod angle to more than $\frac{\pi}{2}$ degrees.

As summarized in Table~\ref{tab:duration}, when trained with $T=50$, neither of the learned models were successful for any 
values of $\alpha$ and $\theta$. However, by increasing the 
training duration $T$, the performance of both designed attack generator models improved. Specifically, for $T=170$ both learned models were able to drive the pendulum rod 
{to fall} without being detected. Also, 
only for shorter training durations~$T$, RNN slightly outperformed FNN, since the RNN was better than FNN in capturing the optimal non-linear  attack dynamics in parts of the state-space not explored during the training. 

We also generated stealthy attacks on both sensors using the model-based LTI method from~\cite{mo2010false,jovanov_tac19,kwon2014analysis} that employs the linearized model of the system dynamics around the zero equilibrium point.
Attacks generated using the LTI model were only
successful for smaller errors -- i.e., $\alpha<4$ and $\theta<\frac{\pi}{3}$. This shows that the \textbf{\textit{LTI-based attacks are only effective around the equilibrium point, where the linearization error is small}}. 


\begin{table}[!t]
    \centering
    \caption{Attack success rate (SR) for different values of training duration $T$ with $N=200$, $\lambda=.5$ and different values of $\alpha$ and $y$ over 100 experiments for the ADV.}
    \vspace{-4pt}
    \setlength\tabcolsep{1.5pt}
    \begin{tabular}{c|c|c|c|c|c|c|c|c}
    \hline
    \makecell{\footnotesize	 SR \% \\ \footnotesize	 FNN/RNN} & \footnotesize	 $|y|=.2$ &  \footnotesize	 $|y|=1$  & \footnotesize	  $|y|=2$ & \footnotesize	  $|y|=4$ & \footnotesize	  $|y|=6$ & \footnotesize	  $|y|=8$ & \footnotesize	  $|y|=10$ & \footnotesize	  $|y|=14$ 
     \\
    \hline
     \footnotesize	 $T=100$    & \footnotesize	 100/100 & \footnotesize	 {0/0}	& \footnotesize	 {0/0} & \footnotesize	 0/0	& \footnotesize	 {0/0} & \footnotesize	 0/0 & \footnotesize	 0/0 & \footnotesize	 0/0  \\
      \footnotesize	 $T=200$    & \footnotesize	{100/100} & \footnotesize	 98/100	& \footnotesize	{16/0} & \footnotesize	 0/0	& \footnotesize	{0/0} & \footnotesize	 0/0 & \footnotesize	 0/0 & \footnotesize	 0/0 \\
     \footnotesize	 $T=300$    & \footnotesize	100/100 & \footnotesize	{100/100}	& \footnotesize	100/100 & \footnotesize	{32/100}	& \footnotesize	{0/0} & \footnotesize	 0/0 & \footnotesize	 0/0 & \footnotesize	 0/0 \\
     \footnotesize	 $T=500$    & \footnotesize	 100/100 & \footnotesize	{100/100} & \footnotesize	{100/100} & \footnotesize	 100/99& \footnotesize	{100/99} & \footnotesize	 50/96 & \footnotesize	 23/96 & \footnotesize	 0/0 \\
     \footnotesize	 $T=700$    & \footnotesize	 100/100 & \footnotesize	{100/100} & \footnotesize	{100/100} & \footnotesize	 100/100& \footnotesize	{100/100} & \footnotesize	 100/100 & \footnotesize	 100/100 & \footnotesize	 99/82 \\
    \hline\hline
    \end{tabular}
    \label{tab:duration_AV}
\end{table}

Table 
\ref{tab:MC_sample} shows the effect of the MC sample number on the performance of the learned attack-generators  for a fixed $T=150$. For smaller values of $N$, FNN performed better; however, when both models had enough data (e.g., $N\geq 250$), both worked equally well. Similarly, Table~\ref{tab:lambda} shows how different values of $\lambda$ in training phase affect the attack model performance. 
For smaller values of $\lambda$, the obtained RNN worked poorly because the data in previous time steps were important to train the model, and {accurately capture the desired attack dynamics}. On the other hand, the learned FNN generator performed worse for very large values~of~$\lambda$.

\subsection{Autonomous Driving Vehicles}


\begin{figure*}[!t]
\begin{subfigure}{0.33\textwidth}
\includegraphics[width=1\linewidth, height=3.55cm]{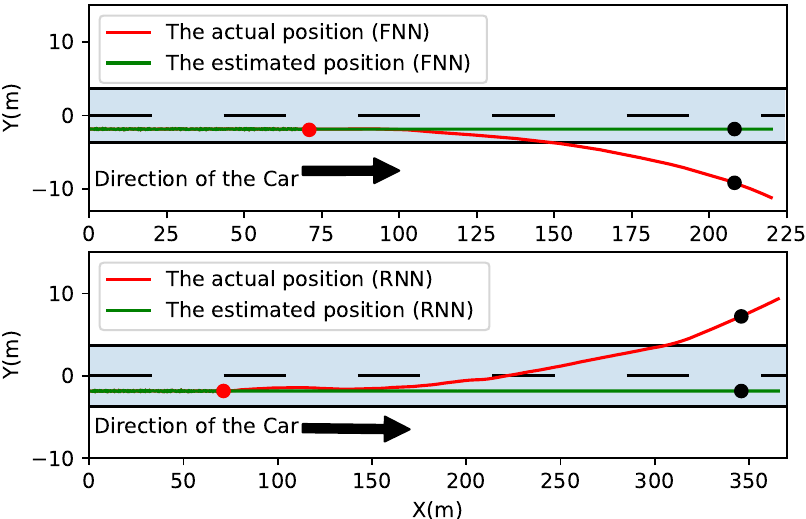}
\label{fig:Trajectory_Straight}
\end{subfigure}
\begin{subfigure}{0.33\textwidth}
\includegraphics[width=1\linewidth, height=3.55cm]{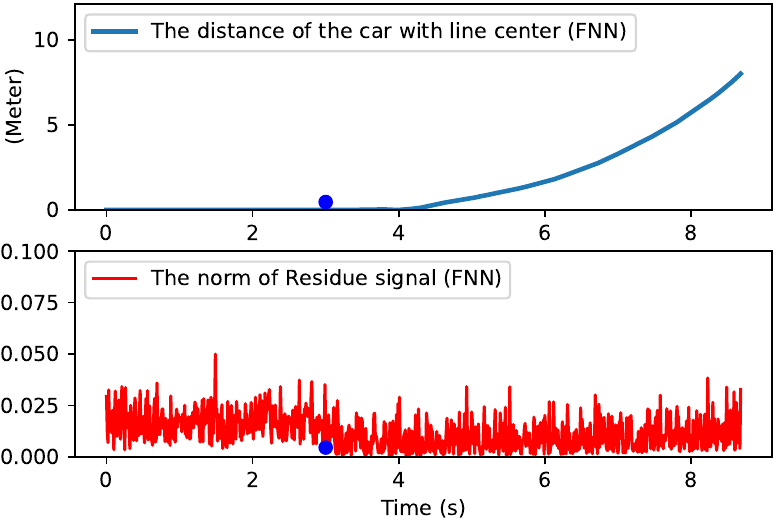}
\label{fig:Error_car_Straight_FNN}
\end{subfigure}
\begin{subfigure}{0.33\textwidth}
\includegraphics[width=1\linewidth, height=3.55cm]{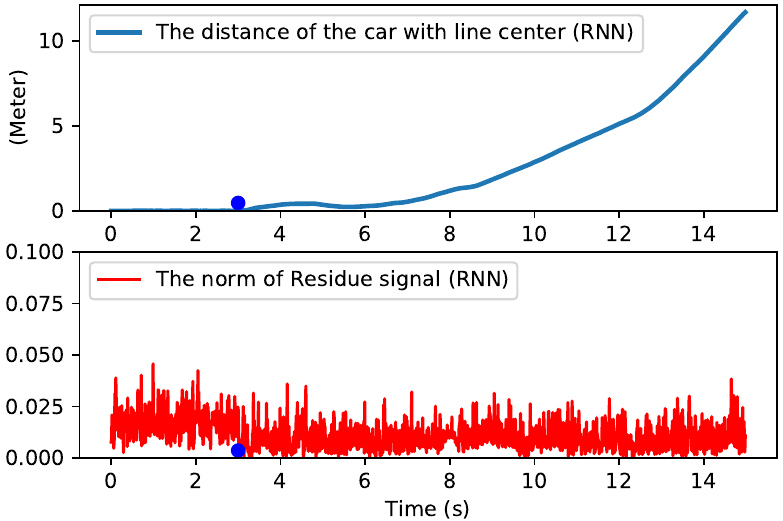}
\label{fig:Error_car_Straight_RNN}
\end{subfigure}
\vspace{-20pt}
\caption{(a)  The trajectory of the compromised car 
(green - the estimated, and the red - the actual vehicle position; the red dot shows the place where the attack started, the black dots show the actual and estimated position of the car at the same time). (b,c) The above sub-figures show the distance of the car with the center of the lane; the bellow sub-figures illustrate the norm of residue signal before and after the start time of attack $t=3s$ (the blue dots) for each of the FNN and RNN-based generators.}
\label{fig:Car1}
\end{figure*} 

\subsubsection{Generic Vehicle Model}
We first considered a simple nonlinear dynamical model of ADV from \cite{kong2015kinematic}, with four states 
$\left[x ~~~y~~~ \psi~~~v\right]^T$;
here, $x$ and $y$ represent the position of the center of mass
in $x$ and $y$ axis, respectively, $\psi$ is the inertial heading, and $v$
is the velocity of the vehicle. 
The states $x, y, \psi$ were measured using noisy sensors, 
with zero-mean noise with covariance matrix $R=.01I$. The system noise was 
zero-mean, with covariance 
$Q=.001I$ and we set the threshold $\eta$ to achieve~$\epsilon=.01$. 


We considered scenario where the car had a constant speed of $25 m/s$, with a feedback controller keeping the car between the lanes. We 
trained offline the FNN and RNN models for generating effective stealthy attacks. The network $H_{\theta}$ was fully connected with 20 neurons and the ReLU activation function, whereas $G_{\theta}$ was an RNN with one layer and 20 neurons, and the ReLU activation~function. 

First, we trained both models with $N=200$ MC samples, $\lambda=.5$ and different duration of training period $T$; we attacked only sensors that measure $y$ and $\psi$ (i.e., not all of the sensors were compromised).  
Table~\ref{tab:duration_AV} shows the success rate (in \%) for both learned attack models for each training period $T$ and different values of $\alpha$ and $|y|$, the car's distance from the center of the lane. As summarized, \textit{increasing the attack training duration helps learn attack generators that can drive the system 
towards the unsafe region} (i.e., increasing $|y|$). 


We also analyzed the impact of $N$, the MC sampling number during training, on attack performance (Table~\ref{tab:MC_sample_AV}); we showed that 
using $N\geq 100$ in training is sufficient. Specifically, 
for smaller values of  $N$, the learned FNN-based generator outperformed the RNN model. However, as $N$ increased both models performed equally good, and even RNN performed slightly better.


\begin{table}[!t]
    \centering
    \caption{Attack success rate (SR) for different values of $N$ with $T=300$, $\lambda=.5$ and different values of $y$ over 100 experiments for the ADV; FNN/RNN  denote FNN vs. RNN SRs.}
     \vspace{-4pt}
    \setlength\tabcolsep{1.5pt}
    \begin{tabular}{c|c|c|c|c|c|c|c|c}
    \hline
    \makecell{\footnotesize	 SR \% \\ \footnotesize	 FNN/RNN} & \footnotesize	 $|y|=.2$ &  \footnotesize	 $|y|=.6$  & \footnotesize	  $|y|=.8$ & \footnotesize	  $|y|=1.2$ & \footnotesize	  $|y|=1.5$ & \footnotesize	  $|y|=2$ & \footnotesize	  $|y|=3$ & \footnotesize	  $|y|=4$ 
     \\
    \hline
     \footnotesize	 $N=1$    & \footnotesize	 100/53 & \footnotesize	 {75/3}	& \footnotesize	 {75/1} & \footnotesize	 75/0	& \footnotesize	 {73/0} & \footnotesize	 0/0 & \footnotesize	 0/0 & \footnotesize	 0/0  \\
      \footnotesize	 $N=10$    & \footnotesize	{100/50} & \footnotesize	 100/41	& \footnotesize	{100/32} & \footnotesize	 100/18	& \footnotesize	{100/12} & \footnotesize	 0/6 & \footnotesize	 0/1 & \footnotesize	 0/0 \\
     \footnotesize	 $N=50$    & \footnotesize	100/99 & \footnotesize	{100/99}	& \footnotesize	100/99 & \footnotesize	{100/99}	& \footnotesize	{100/99} & \footnotesize	 100/99 & \footnotesize	 0/57 & \footnotesize	 0/0 \\
     \footnotesize	 $N=100$    & \footnotesize	 100/100 & \footnotesize	{100/100} & \footnotesize	{100/100} & \footnotesize	 100/100& \footnotesize	{100/100} & \footnotesize	 100/100 & \footnotesize	 21/100 & \footnotesize	 0/0 \\
     \footnotesize	 $N=500$    & \footnotesize	 100/100 & \footnotesize	{100/100} & \footnotesize	{100/100} & \footnotesize	 100/100& \footnotesize	{100/100} & \footnotesize	 100/100 & \footnotesize	 50/100 & \footnotesize	 0/100 \\
    \hline\hline
    \end{tabular}
    \label{tab:MC_sample_AV}
\end{table}


\cref{fig:Car1}(a) shows the trajectory of the car. 
Before starting the attack at the location $X=75 m$, the car (blue line) was kept  between the lanes and the estimated trajectory (green line) had a very small estimation error. However, using either attacks derived by the FNN or RNN-based attack generators, 
the car was being pushed off the road while the estimated position showed that the car was still in the road between the lanes. Furthermore, the attacks were stealthy --  the AD could not detect the presence of either of the attacks. 
\cref{fig:Car1}(b),(c) show the estimation error along the $X$ and $Y$ axis for these~scenarios. 
\subsubsection{Autonomous Driving Simulator: Evaluating on CARLA}
To evaluate our methodology on complex, realistic systems, for which we do not know the model of the 
non-linear vehicle dynamics, we used ADV scenarios in  
vehicle simulator CARLA~\cite{dosovitskiy2017carla}. CARLA is an urban driving simulator 
built on Unreal Engine~4, and providing realistic physics and sensor models in complex urban environments with static and dynamic~actors. 
We defined a planning-navigation-control loop that drove the autonomous vehicle, 
leveraging the EKF structure for $\chi^2$ 
anomaly detector; CARLA setup details are presented in Appendix~\ref{sub:D_E_Carla} and the videos of our CARLA experiments are available at~\cite{CARLA}. 

We 
evaluated our FNN and RNN attack-generators for \emph{performance and generalizability}, and compared to the nominal case without attacks. We 
demonstrated how both FNN and RNN-based attack generators were able to drive the vehicle into unsafe situations (e.g., crashes into other cars or static objects) over short times, while remaining undetected. 
%
We highlight here the results when \textit{not all of the sensors were compromised} (i.e., $\mathcal{K}_a\neq\mathcal{S}$) -- i.e., when the FNN and RNN-based attack generators {\emph{were only able to attack GNSS position measurements and to only have knowledge of positions states}} (i.e., no knowledge of velocity or heading). Despite these restrictions, 
both 
attack generators 
produced stealthy attacks that significantly moved the vehicle off-course 
(e.g., resulting in collisions -- see videos at~\cite{CARLA}).

Additionally, we demonstrated the attack generalizability twofold. 
First, we trained the FNN and RNN-based models offline on a simple path (i.e., not the testing path). 
We then tested those same models on the full CARLA environment and paths. 
Second, we demonstrated the proof-of-concept that the learned attack generator models were 
robust to changes in rates of sensor data by training at 100 Hz measurements and testing at 120 Hz measurements, and retaining attack stealthiness and effectiveness. 

\begin{figure*}[!t]
\begin{subfigure}{0.312\textwidth}
\includegraphics[width=.95\linewidth, height=3.55cm]{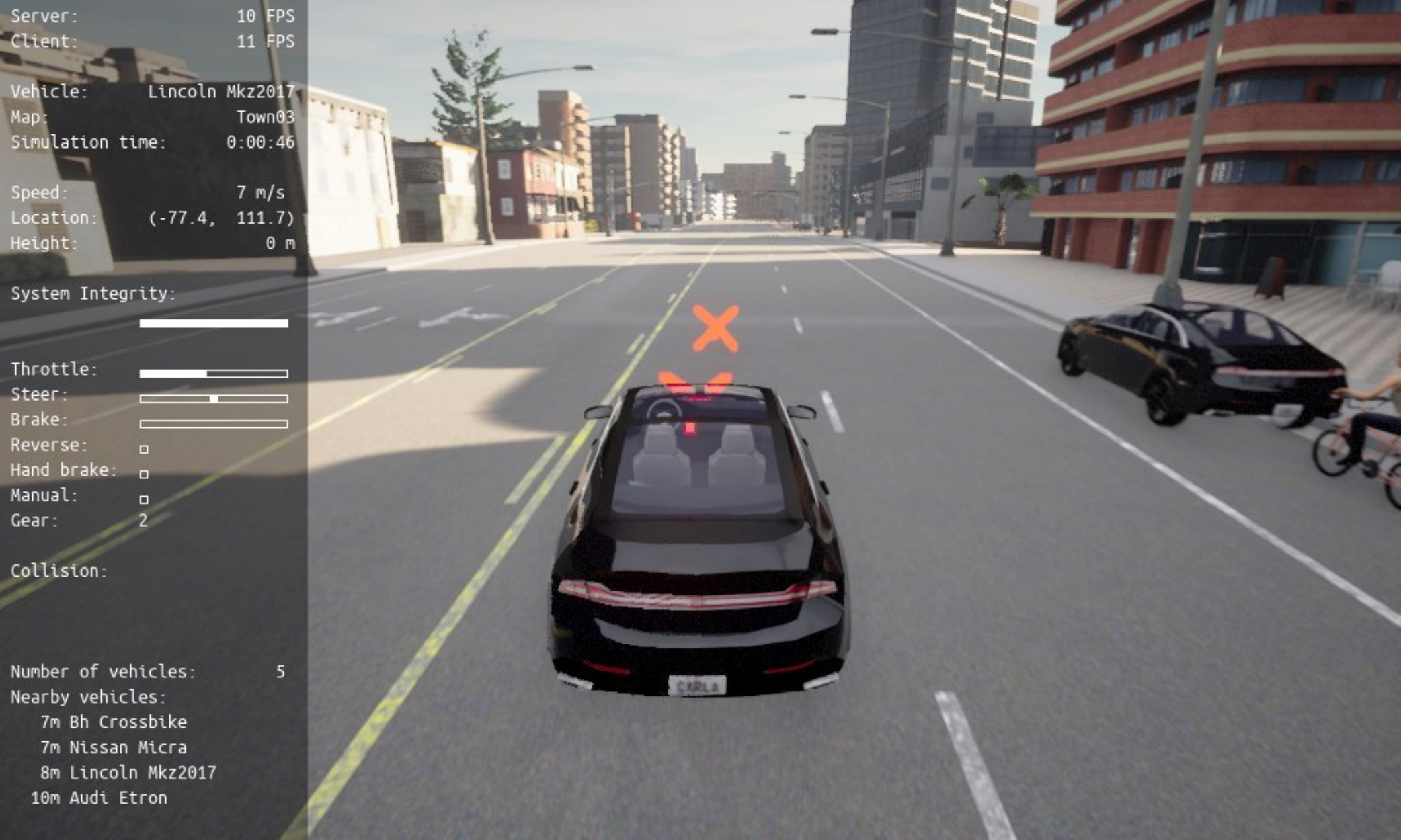}
\label{fig:CARLAa}
\end{subfigure}\hspace{10pt}
\begin{subfigure}{0.312\textwidth}
\includegraphics[width=.95\linewidth, height=3.55cm]{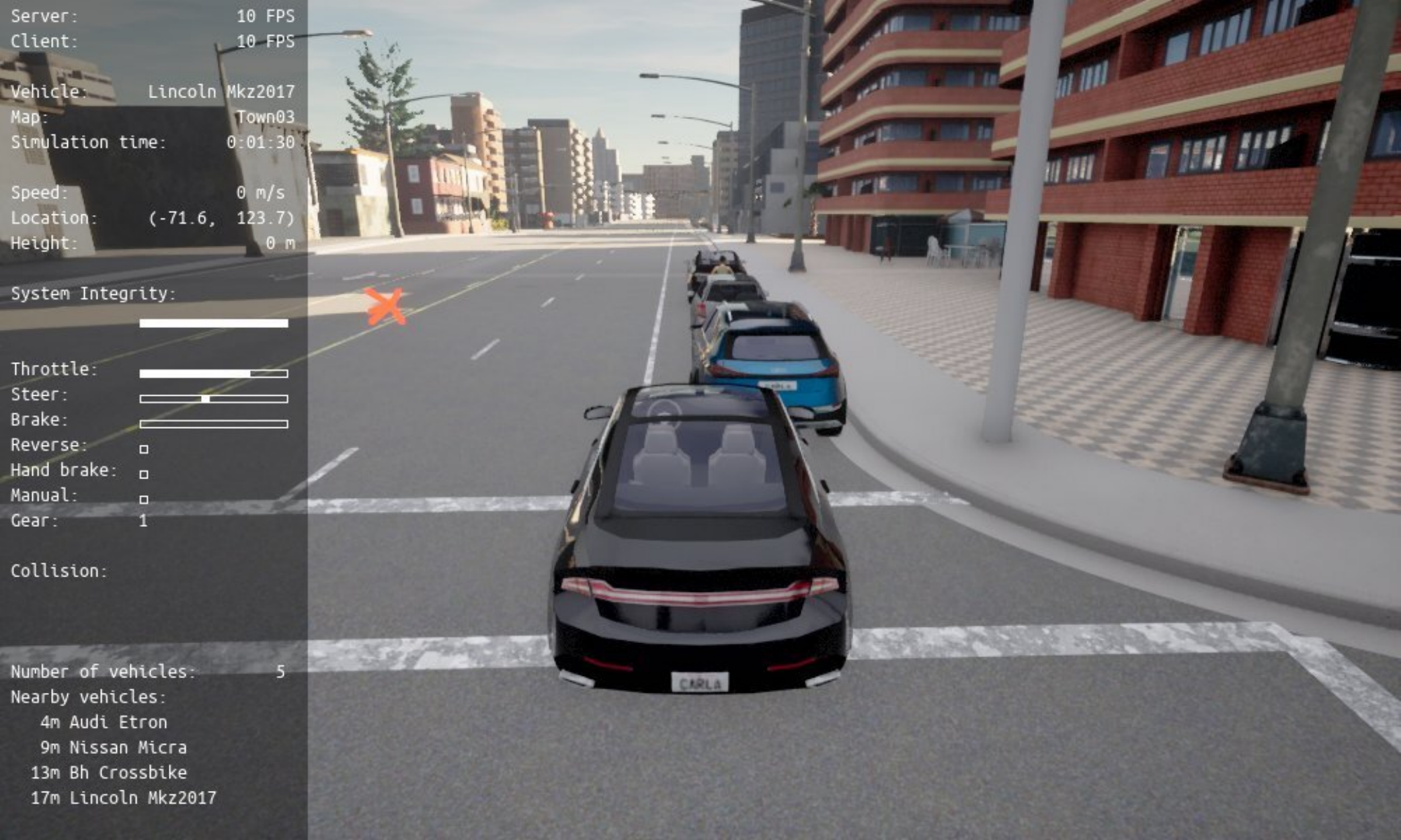}
\label{fig:CARLAb}
\end{subfigure}
\begin{subfigure}{0.312\textwidth}
\includegraphics[width=.95\linewidth, height=3.55cm]{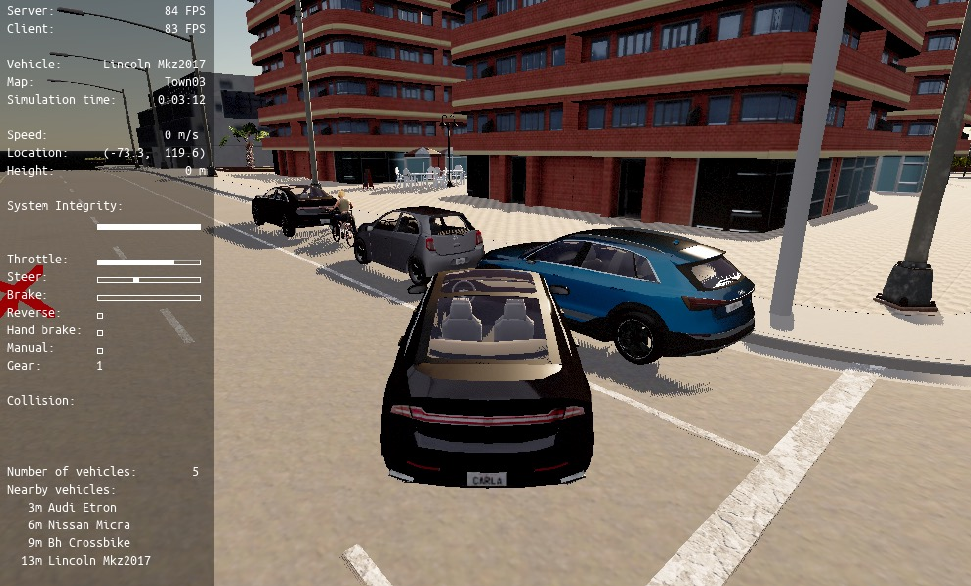}
\label{fig:CARLAc}
\end{subfigure}
\begin{subfigure}{0.312\textwidth}
    \includegraphics[width=.95\linewidth]{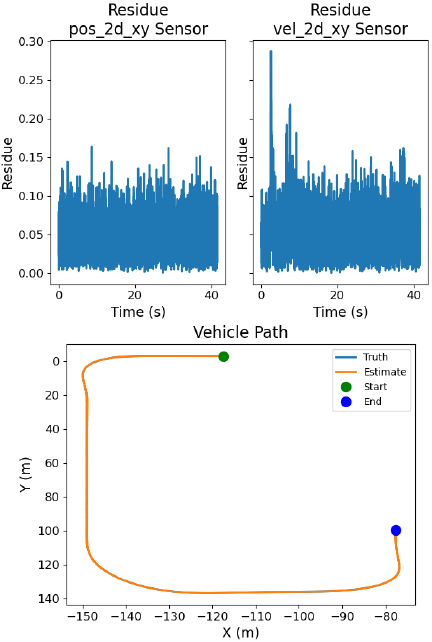}
    \label{fig:CARLAd}
\end{subfigure}
\begin{subfigure}{0.312\textwidth}
    \includegraphics[width=.95\linewidth]{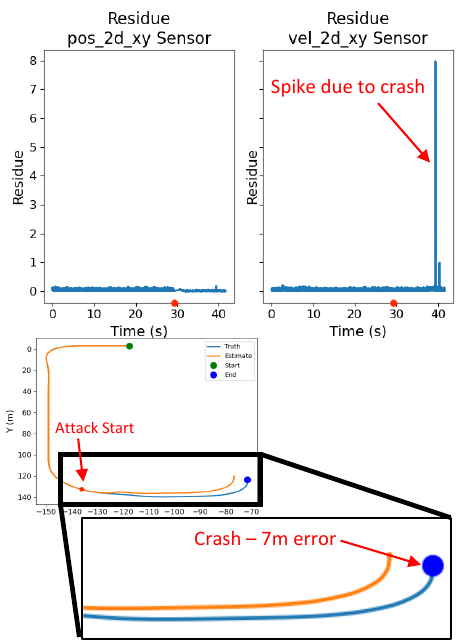}
    \label{fig:CARLAe}
\end{subfigure}
\begin{subfigure}{0.312\textwidth}
    \includegraphics[width=.95\linewidth]{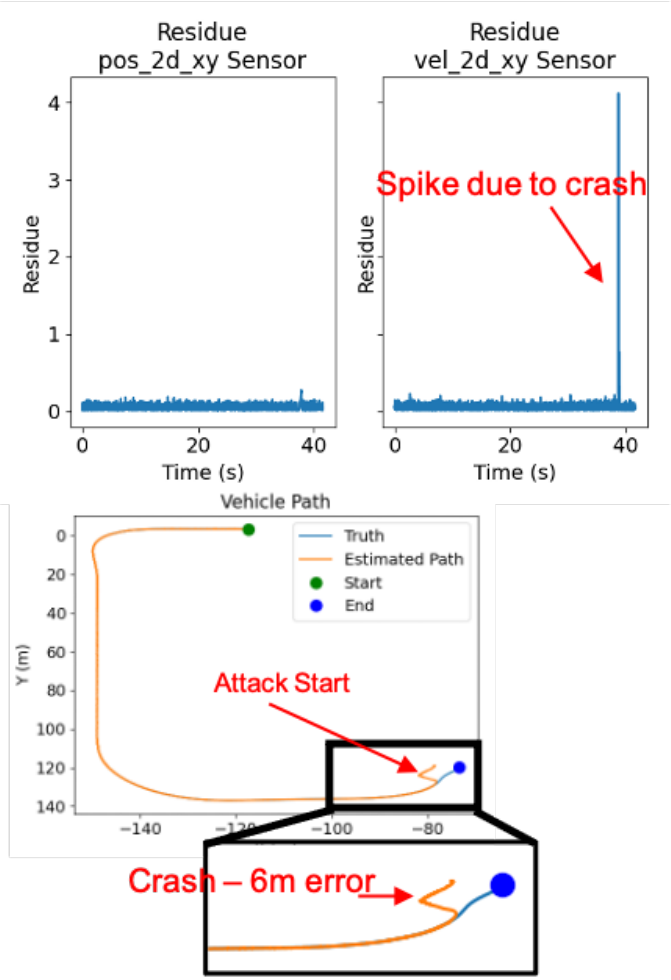}
    \label{fig:CARLAf}
\end{subfigure}
\vspace{-4pt}
\caption{Example results from evaluations on CARLA scenarios: (a)~CARLA simulation 
when the car is free of attack; (b,c)~The vehicle collisions with off-road objects due to the injecting sensor attacks using the FNN and RNN-based attack generators, respectively; (d)~The vehicle trajectory without attack and the residue signals for both velocity and position sensors; (e,f)~The trajectory when the position sensors are compromised using the FNN and RNN-based methods, respectively, and the corresponding residue signals (note different $y-$axes scaling on subfigures (d)-(f). }
\label{fig:CarLA}
\end{figure*}

\cref{fig:CarLA} presents some of the results. Specifically, \cref{fig:CarLA}(d) shows the path and residue values when the sensors were \textbf{not} under attack. \cref{fig:CarLA}(e) and \cref{fig:CarLA}(f) show the path and the residue signal values of the compromised car when the FNN and RNN-based  generators were used to create inserted attack signals. The endpoint of the path is when the car hits an object and stops moving -- the residue signal has a 
spike only when the car hits the object (due to the collision); 
by this point it was too late for any recovery/avoidance~action.


\subsection{Unmanned Aerial Vehicles }
\label{sub:UAV}


Finally, we considered a quadrotor with complex highly nonlinear model from~\cite{bouabdallah2007full} that has 12 states $\begin{bmatrix} x ,\, y ,\, z ,\, \psi ,\, \theta ,\, \phi ,\, \dot{x} , \, \dot{y},\, \dot{z} , \,\dot{\psi} ,\, \dot{\theta}, \, \dot{\phi} \end{bmatrix}^T$;  $x$, $y$ and $z$ represent the quadrotor position along the $X$, $Y$ and $Z$ axis, respectively, while $\dot{x}$, $\dot{y}$ and $\dot{z}$ are their velocity. $\psi$, $\theta$ and $\phi$ are yaw, pitch and roll angles respectively, and $\dot{\psi}$, $\dot{\theta}$ and $\dot{\phi}$ represent their corresponding angular velocity. The system was 
discretized using Euler method with $T_s=.01s$. 
%
The states  $\begin{bmatrix} x ,\, y ,\, z ,\, \psi ,\, \theta ,\, \phi,\dot{\psi} ,\, \dot{\theta}, \, \dot{\phi} \end{bmatrix}^T$ were measured and were affected by zero-mean Gaussian noise with the covariance matrix~$R=.01 I$. We assumed standard disturbance on the input modeled by system noise with zero mean Gaussian with the covariance matrix $Q=.001 I$. We also set $\eta$ such that we obtained $\epsilon=.01$. 
We considered the position control task~\cite{bouabdallah2007full}, where 
the drone should reach a predefined height ($10 m$) and stay there -- i.e., stay at coordinates $X=0$,  $Y=0$ and $Z=10$ if the initial point was denoted as $(0,0,0)$. 
UAV control for this task was based on a standard feedback-based controller. 

\begin{table}[!t]
    \centering
    \caption{Attack success rate (SR) for different values of $T$ with $N=300$, $\lambda=.5$ and different values of $\alpha$ over 100 experiments for the UAV.}
    \vspace{-4pt}
    \setlength\tabcolsep{1.5pt}
    \begin{tabular}{c|c|c|c|c|c|c|c|c}
    \hline
    \makecell{\footnotesize	 SR \% \\ \footnotesize	 FNN/RNN} & \footnotesize	 $\alpha=.2$ &  \footnotesize	 $\alpha=.5$  & \footnotesize	  $\alpha=.7$ & \footnotesize	  $\alpha=1$ & \footnotesize	  $\alpha=3$ & \footnotesize	  $\alpha=5$ & \footnotesize	  $\alpha=7$ & \footnotesize	  $\alpha=9$ 
     \\
    \hline
     \footnotesize	 $T=100$    & \footnotesize	 100/100 & \footnotesize	 {100/87}	& \footnotesize	 {89/49} & \footnotesize	 0/6	& \footnotesize	 {0/0} & \footnotesize	 0/0 & \footnotesize	 0/0 & \footnotesize	 0/0  \\
      \footnotesize	 $T=200$    & \footnotesize	{100/100} & \footnotesize	 95/58	& \footnotesize	{65/53} & \footnotesize	 32/7	& \footnotesize	{0/0} & \footnotesize	 0/0 & \footnotesize	 0/0 & \footnotesize	 0/0 \\
     \footnotesize	 $T=400$    & \footnotesize	100/100 & \footnotesize	{100/100}	& \footnotesize	100/100 & \footnotesize	{67/100}	& \footnotesize	{0/21} & \footnotesize	 0/0 & \footnotesize	 0/0 & \footnotesize	 0/0 \\
     \footnotesize	 $T=600$    & \footnotesize	 100/100 & \footnotesize	{100/100} & \footnotesize	{100/100} & \footnotesize	 100/100& \footnotesize	{80/63} & \footnotesize	 4/9 & \footnotesize	 0/0 & \footnotesize	 0/0 \\
     \footnotesize	 $T=800$    & \footnotesize	 100/100 & \footnotesize	{100/93} & \footnotesize	{100/91} & \footnotesize	 100/91& \footnotesize	{99/83} & \footnotesize	 86/68 & \footnotesize	 69/57 & \footnotesize	 53/43 \\
    \hline 
    \makecell{ \footnotesize	 SR \% \\LTI Model} & \footnotesize	 100 & \footnotesize	 62 & \footnotesize	 0 & \footnotesize	 0 & \footnotesize	 0 & \footnotesize	 0 & \footnotesize	 0 & \footnotesize	 0 
     \\
    \hline\hline
    \end{tabular}
    \label{tab:T_UAV_alpha}
\end{table}

\begin{table}[!t]
    \centering
    \caption{Attack success rate (SR) for different values of $T$ with $N=300$, $\lambda=.5$ and different values of $x$ along $X$ axis over 100 experiments for the UAV.}
    \vspace{-4pt}
    \setlength\tabcolsep{1.5pt}
    \begin{tabular}{c|c|c|c|c|c|c|c|c}
    \hline
    \makecell{\footnotesize	 SR \% \\ \footnotesize	 FNN/RNN} & \footnotesize	 $|x|=.2$ &  \footnotesize	 $|x|=.4$  & \footnotesize	  $|x|=.6$ & \footnotesize	  $|x|=.8$ & \footnotesize	  $|x|=1$ & \footnotesize	  $|x|=1.5$ & \footnotesize	  $|x|=3$ & \footnotesize	  $|x|=4$ 
     \\
    \hline
     \footnotesize	 $T=100$    & \footnotesize	 93/50 & \footnotesize	 {1/2}	& \footnotesize	 {0/0} & \footnotesize	 0/0	& \footnotesize	 {0/0} & \footnotesize	 0/0 & \footnotesize	 0/0 & \footnotesize	 0/0  \\
      \footnotesize	 $T=200$    & \footnotesize	{47/5} & \footnotesize	 17/0	& \footnotesize	{5/0} & \footnotesize	 1/0	& \footnotesize	{0/0} & \footnotesize	 0/0 & \footnotesize	 0/0 & \footnotesize	 0/0 \\
     \footnotesize	 $T=400$    & \footnotesize	100/57 & \footnotesize	{44/14}	& \footnotesize	0/5 & \footnotesize	{0/0}	& \footnotesize	{0/0} & \footnotesize	 0/0 & \footnotesize	 0/0 & \footnotesize	 0/0 \\
     \footnotesize	 $T=600$    & \footnotesize	 83/78 & \footnotesize	{66/24} & \footnotesize	{50/4} & \footnotesize	 37/2& \footnotesize	{30/1} & \footnotesize	 13/0 & \footnotesize	 0/0 & \footnotesize	 0/0 \\
     \footnotesize	 $T=800$    & \footnotesize	 97/77 & \footnotesize	{90/36} & \footnotesize	{85/22} & \footnotesize	 75/13& \footnotesize	{68/10} & \footnotesize	 51/1 & \footnotesize	 0/0 & \footnotesize	 0/0 \\
    \hline\hline
    \end{tabular}
    \label{tab:T_UAV_x}
\end{table}

$G_{\theta}$ used for synthesizing the stealthy attack was a 2-layer RNN with ReLU activation function and 55 neurons per layer. $H_{\theta}$ was also a 2-layer 
FNN with ReLU activation function and $55$ neuron for each layer. First, we trained both models with $N=300$ MC samples, $\delta=.2$, $\lambda=.5$ and different values of training period $T$. {We also considered the case where all sensor are under attack.}  

Tables~\ref{tab:T_UAV_alpha}-\ref{tab:T_UAV_z} (due to space constraints, Tables~\ref{tab:T_UAV_y} and \ref{tab:T_UAV_z} are in Appendix~\ref{sub:ASR}) 
show the success rate (in \%) for both learned attack generators obtained for each training period $T$ and different values of $\alpha$, $|x|$, $|y|$ and $|z|$ (i.e., the drone's distance from the desired position along each axis). As summarized, increasing the attack training duration helps learn effective stealthy attack generators capable of
driving the system 
towards the unsafe region (i.e., increasing $|x|$, $|y|$ and $|z|$). 
Furthermore, for suitably large training periods $T$, on average the learned FNN-based attack generators outperforms the RNN-based attack generators. 
{Moreover, we evaluated the effectiveness of the LTI-based attacks (i.e., which linearize the UAV model)~\cite{mo2010false,jovanov_tac19,kwon2014analysis}; our results, summarized in the last line in Table~\ref{tab:T_UAV_alpha}, show 
that the LTI-based attacks are only 62\% successful in reaching $\alpha=.2$, and unsuccessful for higher values of $\alpha$, for the same reasons as in the pendulum study -- linearization error becomes too large for large state deviations, limiting their~applicability.}

\begin{figure*}[!t]
\begin{subfigure}{0.342\textwidth}
\includegraphics[width=1.1\linewidth]{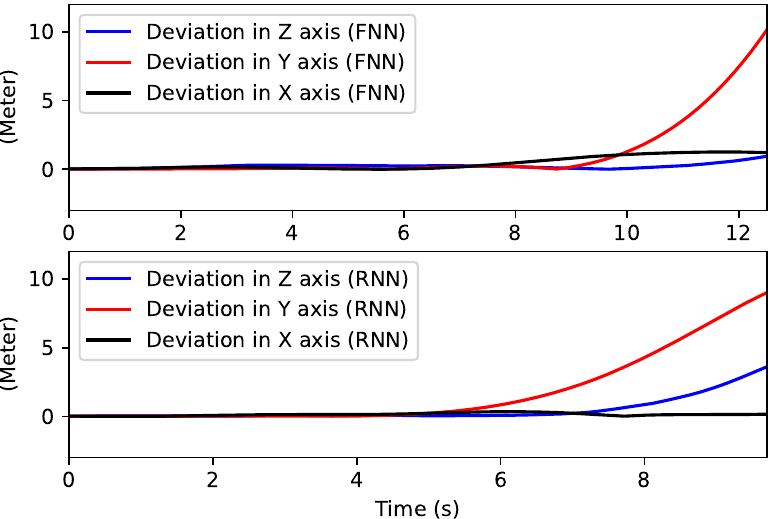} 
\label{fig:uav_a}
\end{subfigure}
\hspace{64pt}
\begin{subfigure}{0.342\textwidth}
\includegraphics[width=1.1\linewidth]{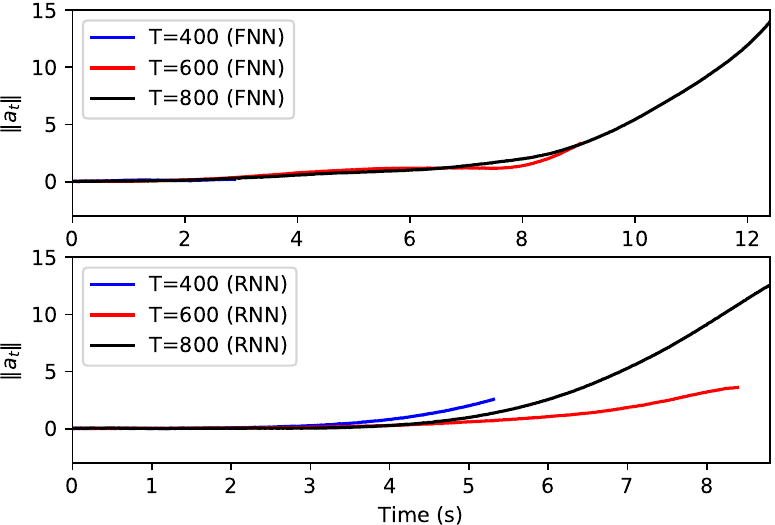}
\label{fig:uav_b}
\end{subfigure}
\vspace{-12pt}
\caption{UAV altitude control: (a)  The deviation of the drone along each axis from the desired coordinates $X=0$,  $Y=0$ and $Z=10$, for RNN and FNN-based attack (the attack starts at time zero);  (b) The norm of the attack vector over time for both FNN and RNN attack generators with $T=400,600, 800$.} 
\label{fig:uav}
\end{figure*}

\cref{fig:uav}(a) illustrates the deviation of the drone from the desired hovering point over time for a successful attack sequence obtained from a generator trained with $T=800$ and $N=300$. The attack started at $t=0$; 
over time, the drone's deviation from the desired position will increase. \cref{fig:uav}(b) shows the norm of the attack vector of both FNN and RNN models for three different attack-generator models trained with $T=200, 400$ and $600$. Note that unlike adversarial machine learning in other domains (e.g., image classification) where the norm of the attack is limited to be bounded, the stealthiness condition in CPS requires the norm of the attack vector to gradually increase over~time.

\section{Conclusion}
\label{sec:concl}

In this work, we have utilized deep learning to generate stealthy attacks on control components in cyber-physical systems, focusing on a widely used architecture 
where the low-level control employs the extended Kalman filter 
and an anomaly detector.
We have considered a grey box setup,
with unknown nonlinear plant dynamics and known observation functions and Kalman filter gains.
We have shown that feedforward and recurrent neural networks (FNN and RNN, respectively) can be
used to generate stealthy adversarial attacks on sensing information delivered to the system, 
resulting in large errors to the estimates of the state of the system 
without being detected.
Both FNN and RNN are trained offline from a cost function combining the attack effects on the estimation error and the residual signal of the EKF; thus, the trained model is capable of recursively generating such effective sensor attacks in real-time using only current sensor measurements. 
The effectiveness of the proposed methods has been illustrated and evaluated on several case studies with varying complexity. 

\bibliographystyle{ACM-Reference-Format}

\section{Appendix}
\label{app:app}

\subsection{Proof of Theorem~\ref{Thm:PA}}
\label{app:T1proof}

First, we will show that applying such attack sequence results in an unbounded estimation error. 
For LTI systems, 
the dynamic of the state estimation error follows
\begin{equation*}
\begin{split}
\Delta{x}_{t}&=x_t-\hat{x}_t\\
&=A\Delta{x}_{t-1}+w_t-L(y^c_t-CA\hat{x}_{t-1}-CBu_{t-1})\\
&=A\Delta{x}_{t-1}+w_t-L\phi_t
\end{split}
\end{equation*}
As the matrix $A$ is unstable, it 
follows that 
$\|\Delta x_t\|$ will be unbounded as $t \to \infty$. 

We now show that the attack is stealthy from the perspective of the 
IDS. In this case, the residue signal $z_t$ satisfies
%
\begin{equation}
\begin{split}
z_t&=y^c_t-C(A\hat{x}_{t-1}+Bu_{t-1})=\\
&=y_t+a_t-C(A\hat{x}_{t-1}+Bu_{t-1})=\phi_t.
\end{split}
\end{equation}
Therefore, it follows that
\begin{equation*}
\begin{split}
\mathbb{E}\{g_t^a\}&=\mathbb{E}\{z_t^TS^{-1}
z_t\}=\mathbb{E}\{\phi_t^TS^{-1} \phi_t\}\\&=trace(\mathbb{E}\{\phi_t^TS^{-1} \phi_t\})
=\mathbb{E}\{trace(\phi_t \phi_t^TS^{-1})\}\\&=trace(\mathbb{E}\{\phi_t\phi_t^T\}S^{-1})
\leq trace(SS^{-1})=p,
\end{split}
\end{equation*}
where we used the linearity of expectation and trace operation. Note that for LTI systems, the expectation of $g_t$  (also known as the degrees of freedom of the distribution) satisfies that $\mathbb{E}\{g_t\}=p$. Based on the properties of the $\chi^2$ distribution, since  $\mathbb{E}\{g_t^a\} \leq \mathbb{E}\{g_t\}=p$, it follows that $\mathbb{P}(g_t^a>\eta)\leq \mathbb{P}(g_t>\eta)=\epsilon$, and thus the attack sequence is~stealthy.

\subsection{Proof of Theorem~\ref{thm:bound}}
\label{app:T2proof}

{From the multivariate Chebyshev’s inequality \cite{navarro2016very}, it holds that $\mathbb{P}( v_t^T R^{-1}v_t \leq k^2)\geq 1-\frac{p}{k^2}$. On the other hand, using our assumption $R\preceq \sigma I$, it holds that $\sigma^{-1}v_t^Tv_t\leq v_t^T R^{-1}v_t$ for any $v_t\in \mathbb{R}^{p}$. Therefore, $\mathbb{P}( v_t^T v_t \leq \sigma k^2)\geq 1-\frac{p}{k^2}$ or equivalently $\mathbb{P}( \Vert v_t\Vert \leq \sqrt{\sigma} k)\geq 1-\frac{p}{k^2}$.} Now, with the probability of at least $1-\frac{p}{k^2}$, we have that 
\begin{equation*}
\begin{split}
\alpha &\leq  \Vert y_t-h(\hat{x}_t) \Vert = \Vert h(x_t)+v_t-h(\hat{x}_t) \Vert \leq \\
&\leq L \Vert x_t-\hat{x}_t \Vert + \Vert v_t \Vert\leq L \Vert x_t-\hat{x}_t \Vert + \sqrt{\sigma} k,
\end{split}
\end{equation*}
which results in $ \Vert x_t-\hat{x}_t \Vert \geq \frac{\alpha - \sqrt{\sigma}k}{L}$.

\subsection{Details of Employed CARLA Setup}\label{sub:D_E_Carla}

For planning, CARLA provides with a state-machine waypoint following algorithm. A vehicle’s (estimated) pose and velocity were used along with map-based waypoints to coordinate (i)~road-following, (ii)~left-turn, (iii)~right-turn, (iv)~intersection, and (v)~hazard-stop conditions~\cite{dosovitskiy2017carla}. We estimated the pose and velocity using an EKF with high-rate sensor data. 

We also leverage the EKF structure to design an industry-standard 
$\chi^2$ anomaly detector (AD). 
We set threshold $\eta$ to result in $\epsilon=.05$ in normal condition. Then, the integrity value shown in the left bar of 
\cref{fig:CarLA}(a),(b),(c) represents the number of measurements that pass the $\chi^2$ AD requirement out of the last 20 measurements. We assume that the attack is detected if more than two sensor measurements cannot pass the requirement in this window of  time.
As sensor inputs, a Global Navigation Satellite Sensor (GNSS) sensor provides loosely coupled position solutions in global coordinates, a commercial GNSS standard. We similarly define a generalized velocimeter model, derived from Doppler (or, more frequently in safety-critical applications, GNSS delta-range). 

State estimates and planning objectives were feed into a standard feedback controller~\cite{emirler2014robust} that targeted a cruising speed of 25 km/hr ($\sim 7 m/s$). The control algorithm drove the following actuators with associated input ranges: (i) Steering wheel angle on $[-1.0, 1.0]$, (ii) Throttle on $[0.0, 1.0]$, and (iii) Brake on $[0.0, 1.0]$.
Finally, we visualize the vehicle trajectory and system integrity with a heads-up-display presented in \cref{fig:CarLA}; the videos for our CARLA experiments are available at~\cite{CARLA}.

\subsection{Analysis of Attack Success Rate on Unmanned Aerial Vehicles}
\label{sub:ASR}

Tables~\ref{tab:T_UAV_y} and \ref{tab:T_UAV_z} summarize the attack success rate (ASR) for both learned attack generators for UAVs; the generators were obtained for each training period $T$ and different values of $|y|$ and $|z|$ (i.e., the drone's distance from the desired position along these two axis).

\begin{table}[h]
    \centering
    \caption{Attack success rate (SR) for different values of $T$ with $N=300$, $\lambda=.5$ and different values of $y$ along $Y$ axis over 100 experiments for the UAV.}
    \vspace{-4pt}
    \setlength\tabcolsep{1.5pt}
    \begin{tabular}{c|c|c|c|c|c|c|c|c}
    \hline
    \makecell{\footnotesize	 SR \% \\ \footnotesize	 FNN/RNN} & \footnotesize	 $|y|=.2$ &  \footnotesize	 $|y|=.5$  & \footnotesize	  $|y|=.8$ & \footnotesize	  $|y|=1$ & \footnotesize	  $|y|=2$ & \footnotesize	  $|y|=4$ & \footnotesize	  $|y|=6$ & \footnotesize	  $|y|=8$ 
     \\
    \hline
     \footnotesize	 $T=100$    & \footnotesize	 0/44 & \footnotesize	 {0/0}	& \footnotesize	 {0/0} & \footnotesize	 0/0	& \footnotesize	 {0/0} & \footnotesize	 0/0 & \footnotesize	 0/0 & \footnotesize	 0/0  \\
      \footnotesize	 $T=200$    & \footnotesize	{92/91} & \footnotesize	 24/20	& \footnotesize	{0/0} & \footnotesize	 0/0	& \footnotesize	{0/0} & \footnotesize	 0/0 & \footnotesize	 0/0 & \footnotesize	 0/0 \\
     \footnotesize	 $T=400$    & \footnotesize	2/100 & \footnotesize	{0/100}	& \footnotesize	0/96 & \footnotesize	{0/93}	& \footnotesize	{0/25} & \footnotesize	 0/0 & \footnotesize	 0/0 & \footnotesize	 0/0 \\
     \footnotesize	 $T=600$    & \footnotesize	 99/100 & \footnotesize	{94/97} & \footnotesize	{92/95} & \footnotesize	 91/94& \footnotesize	{73/66} & \footnotesize	 0/0 & \footnotesize	 0/0 & \footnotesize	 0/0 \\
     \footnotesize	 $T=800$    & \footnotesize	 100/91 & \footnotesize	{98/91} & \footnotesize	{96/91} & \footnotesize	 95/90& \footnotesize	{88/81} & \footnotesize	 73/65 & \footnotesize	 62/51 & \footnotesize	 45/41 \\
    \hline\hline
    \end{tabular}
    \label{tab:T_UAV_y}
\end{table}

\begin{table}[h]
    \centering
    \caption{Attack success rate (SR) for different values of $T$ with $N=300$, $\lambda=.5$ and different values of $z$ along $Z$ axis over 100 experiments for the UAV.}
    \vspace{-4pt}
    \setlength\tabcolsep{1.5pt}
    \begin{tabular}{c|c|c|c|c|c|c|c|c}
    \hline
    \makecell{\footnotesize	 SR \% \\ \footnotesize	 FNN/RNN} & \footnotesize	 $|z|=.2$ &  \footnotesize	 $|z|=.4$  & \footnotesize	  $|z|=.6$ & \footnotesize	  $|z|=.8$ & \footnotesize	  $|y|=1$ & \footnotesize	  $|z|=2$ & \footnotesize	  $|z|=3$ & \footnotesize	  $|z|=4$ 
     \\
    \hline
     \footnotesize	 $T=100$    & \footnotesize	 0/9 & \footnotesize	 {0/0}	& \footnotesize	 {0/0} & \footnotesize	 0/0	& \footnotesize	 {0/0} & \footnotesize	 0/0 & \footnotesize	 0/0 & \footnotesize	 0/0  \\
      \footnotesize	 $T=200$    & \footnotesize	{67/0} & \footnotesize	 24/0	& \footnotesize	{4/0} & \footnotesize	 0/0	& \footnotesize	{0/0} & \footnotesize	 0/0 & \footnotesize	 0/0 & \footnotesize	 0/0 \\
     \footnotesize	 $T=400$    & \footnotesize	44/42 & \footnotesize	{0/17}	& \footnotesize	0/10 & \footnotesize	{0/7}	& \footnotesize	{0/4} & \footnotesize	 0/2 & \footnotesize	 0/0 & \footnotesize	 0/0 \\
     \footnotesize	 $T=600$    & \footnotesize	 86/81 & \footnotesize	{59/74} & \footnotesize	{45/60} & \footnotesize	 26/48& \footnotesize	{11/36} & \footnotesize	 3/28 & \footnotesize	 0/3 & \footnotesize	 0/0 \\
     \footnotesize	 $T=800$    & \footnotesize	 95/88 & \footnotesize	{88/85} & \footnotesize	{78/83} & \footnotesize	 69/80& \footnotesize	{64/75} & \footnotesize	 40/66 & \footnotesize	 0/41 & \footnotesize	 0/6 \\
    \hline\hline
    \end{tabular}
    \label{tab:T_UAV_z}
\end{table}

\end{document}